\documentclass[aps,prd,superscriptaddress,twocolumn,notitlepage,nofootinbib]{revtex4-2}
\usepackage{aas_macros}
\usepackage[normalem]{ulem}
\usepackage[utf8]{inputenc}
\usepackage[english]{babel}
\usepackage{amsmath}
\usepackage{amsfonts}
\usepackage{amssymb}
\usepackage{listings}
\usepackage{lipsum}
\usepackage{multirow}
\usepackage{datetime}
\usepackage{graphicx}
\usepackage{mathtools}
\usepackage{mathrsfs}
\usepackage{dcolumn}
\usepackage{comment}
\usepackage{multirow}
\usepackage{physics}
\usepackage{orcidlink}
\usepackage{longtable}
\usepackage{afterpage} 
\usepackage{ragged2e}   
\usepackage{soul}
\usepackage{color}   
\usepackage{xcolor}

\usepackage{hyperref}
\hypersetup{
    colorlinks=true, 
    pdfborder = {0 0 0.5 [3 3]},
    anchorcolor=black,
    citecolor=blue,
    linktoc=all,    
    linktocpage=true,
    linkcolor=red,
	urlcolor=blue
}

\newcommand{\M}{M}
\newcommand{\dpm}{m}
\newcommand{\Tfft}{T_{\rm FFT}}

\newcommand{\Perimeter}{Perimeter Institute for Theoretical Physics, Waterloo, Ontario N2L 2Y5, Canada}
\newcommand{\UdSCag}{Physics Department, Universit\`a degli Studi di Cagliari, Cagliari  09042, Italy}
\newcommand{\INFNCA}{INFN, Sezione di Cagliari, Cagliari 09042, Italy}
\newcommand{\INFNRM}{INFN, Sezione di Roma, Piazzale Aldo Moro, 2, I-00185 Roma, Italy}
\newcommand{\Sapienza}{Physics Department, Universit\`a di Roma “Sapienza”, Piazzale Aldo Moro, 2, I-00185 Roma, Italy}
\newcommand{\PGI}{Princeton Gravity Initiative, Princeton University, Princeton New Jersey 08544, USA}
\newcommand{\Princeton}{Department of Physics, Princeton University, Princeton, New Jersey 08544, USA}

\graphicspath{{figs/}}

\makeatletter
\def\l@subsubsection#1#2{}
\makeatother

\begin{document}

\title{Search for continuous gravitational wave signals from luminous dark photon superradiance clouds with LVK O3 observations}

\author{Lorenzo Mirasola\,\orcidlink{0009-0004-0174-1377}}\email[]{lorenzo.mirasola@ca.infn.it}\affiliation{\UdSCag}\affiliation{\INFNCA}
\author{Cristina Mondino\,\orcidlink{0000-0002-8058-4055}}\email[]{cmondino@perimeterinstitute.ca}\affiliation{\Perimeter}
\author{Francesco Amicucci\orcidlink{0009-0005-2139-4197}}\affiliation{\Sapienza}\affiliation{\INFNRM}
\author{Nils Siemonsen\orcidlink{0000-0001-5664-3521}}\affiliation{\PGI}\affiliation{\Princeton}
\author{Cristiano Palomba\,\orcidlink{0000-0002-4450-9883}}\affiliation{\INFNRM}
\author{Sabrina D'Antonio\,\orcidlink{0000-0003-0898-6030}}\affiliation{\INFNRM}
\author{Paola Leaci\,\orcidlink{0000-0002-3997-5046}}\affiliation{\Sapienza}\affiliation{\INFNRM}
\author{Luca D'Onofrio\,\orcidlink{0000-0001-9546-5959}}\affiliation{\INFNRM}
\author{Pia Astone\,\orcidlink{0000-0003-4981-4120}}\affiliation{\INFNRM}
\author{Daniel Ega\~na-Ugrinovic\orcidlink{0000-0003-2680-9070}}
\affiliation{\Perimeter}
\author{Junwu Huang\orcidlink{0000-0001-6007-7315}}
\affiliation{\Perimeter}
\author{Masha Baryakhtar\orcidlink{0000-0002-7631-2604}}\affiliation{Physics Department, University of Washington, Seattle, Washington 98195-1560, USA}
\author{William E.\ East\orcidlink{0000-0002-9017-6215}}
\affiliation{\Perimeter}

\date{\today}

\begin{abstract}
Superradiance clouds of kinetically mixed dark photons around spinning black holes can produce observable multimessenger electromagnetic and gravitational wave signals.
The cloud generates electric fields of up to a teravolt-per-meter, which leads to a cascade production of charged particles, yielding a turbulent quasiequilibrium plasma around the black hole, and resulting in electromagnetic fluxes ranging from supernova to pulsar-like luminosities. For stellar mass black holes, such systems resemble millisecond pulsars and are expected to emit pulsating radio waves and continuous gravitational waves (CWs) within the LIGO-Virgo-KAGRA (LVK) sensitivity band. We select 44 sources with approximately coincident frequencies or positive frequency drifts from existing pulsar catalogs as potential candidates of long-lasting superradiance clouds around old Galactic black holes. 
For a subset of 34 sources that are well measured and have not been previously targeted, we perform the first search for CW emission in LVK data from the third observing run. 
We find no evidence of a CW signal and place 95\% confidence level upper limits on the emitted strain amplitude. 
We interpret these results, together with limits from previous searches, in terms of the underlying dark photon theory by performing an analysis of the expected signals from superradiance clouds from Galactic black holes. We find that, even for moderately spinning black holes, the absence of an observed CW signal disfavors a discrete set of dark photon masses between about $10^{-13}$ and $10^{-12}$ $\rm{eV}/c^2$ and kinetic mixing couplings in the range of $10^{-9}$--$10^{-7}$, subject to assumptions about the properties of the black hole population and the cloud's electromagnetic emission.

\end{abstract}

\maketitle
\tableofcontents

\section{Introduction}
The current generation of gravitational wave (GW) detectors has opened a new era in the exploration of astrophysics and fundamental physics. The advanced Laser Interferometer Gravitational Wave Observatory (aLIGO)~\cite{ligo}, Advanced Virgo~\cite{virgo}, and KAGRA~\cite{kagra} have already observed 90 confirmed compact binary coalescence~\cite{KAGRA:2021vkt,gwosc} events in the first three observing runs [O1 + O2~\cite{LIGOScientific:2018mvr} and O3~\cite{LIGOScientific:2020ibl, KAGRA:2021vkt}] and more than a hundred preliminary candidates from the ongoing O4 run~\cite{publicalters}.
The wealth of detections is revolutionizing multiple areas of astrophysics, providing invaluable information about black hole and neutron star properties. Beyond the measurement of transient coalescence events, other types of signals can be searched for with the LIGO-Virgo-KAGRA (LVK) detectors network, including transient GW bursts, long-lasting continuous GWs (CWs), and stochastic GW backgrounds. 

The primary targets of CW observations are rotating neutron stars with asymmetric mass distributions relative to their rotation axes~\cite{1989thyg.book.....H}. In recent years, CW sources due to new physics beyond the Standard Model of particle physics have also been proposed. The search for CWs presents unique challenges in data analysis, primarily due to the expected weakness and persistence of the signals. Specifically, the long duration of CW signals requires considerations that differ from those of the short-duration binary coalescence events that have been detected so far. Notably, the extended duration necessitates explicitly accounting for the motion of the detectors and their time-dependent response to waves originating from a particular direction in the sky. Additionally, relativistic effects in wave propagation must be carefully considered. 
It is expected that the detection of CWs will open a new window in the observation of the GW sky, allowing for long-term monitoring of sources. This will enable very precise measurements of source parameters, reveal subtle effects that accumulate over time, and possibly uncover evidence of new physics. For comprehensive reviews on CW sources and searches, see, e.g.,~\cite{Riles:2022wwz,Piccinni:2022vsd,Wette:2023dom}.

A particularly promising and well-motivated new physics target for GW detectors is given by new ultralight bosonic fields that form gravitationally-bound ``clouds'' around spinning black holes (BHs)~\cite{Arvanitaki:2009fg,Arvanitaki:2010sy} through the superradiance instability~\cite{Zeldovich1971,Misner1972,Starobinskii1973,Detweiler:1980uk,Bekenstein:1998nt, Brito:2015oca}. Independently of the boson's initial abundance, if its Compton wavelength is of the order of the BH size, the cloud grows to a coherent state with an exponentially large occupation number  at the expense of the BH's energy and angular momentum~\cite{Arvanitaki:2009fg,Arvanitaki:2010sy}. The growth stops when a significant fraction of the initial BH spin has been extracted. This phenomenon, which only requires gravitational interactions, occurs for both scalar and vector fields
and leads to three main observational signatures in current GW detectors~\cite{Arvanitaki:2014wva, Arvanitaki:2016qwi, Brito:2017zvb}. The first is an indirect imprint in the spin-mass distribution in merging BH populations, effectively removing old, highly-spinning BHs~\cite{Arvanitaki:2016qwi, Brito:2017zvb, Baryakhtar:2017ngi}. Mass and spin measurements of merging BHs from the LVK catalogs have already been used to disfavour a factor of two in noninteracting scalar mass parameter space~\cite{Ng:2019jsx,Ng:2020ruv}, subject to assumptions on the time between BH formation and the merger. Spin measurements of old, highly spinning BHs in x-ray binaries have also been used to place constraints on ultralight bosons~\cite{Arvanitaki:2014wva, Baryakhtar:2017ngi, Cardoso:2018tly, Baryakhtar:2020gao, Hoof:2024quk}, albeit potentially affected by systematic uncertainties from modeling of the accretion disk. Spin reduction due to superradiance has also been shown to increase the rate of hierarchical BH mergers in nuclear star clusters~\cite{Payne:2021ahy}.

The second observational effect of BH superradiance is a direct signal of quasi-monochromatic and long-lived GWs that are emitted through dissipation of the rotating boson cloud~\cite{Arvanitaki:2010sy,Yoshino:2013ofa}. Such signals can be loud enough to be observed in LVK for clouds around Galactic and extra-Galactic BHs, and the nondetection in all-sky and Galactic Center CW searches has been used to place constraints on the axion mass vs BH mass parameter space~\cite{Palomba:2019vxe, LIGOScientific:2021rnv, KAGRA:2022osp}. All-sky searches have also been interpreted as bounds on the axion mass by estimating the expected observable signals from Galactic BHs, subject to assumptions on the BH population properties~\cite{Zhu:2020tht}. 
Boson clouds can also contribute to the stochastic GW background~\cite{Brito:2017zvb} and corresponding searches have been used to constrain scalar masses with O1, O2 and O3 data~\cite{Tsukada:2018mbp, Yuan:2022bem} and vector masses with O1 and O2 results~\cite{Tsukada:2020lgt}. Further probes of GW signals from boson clouds include a directed search targeting the BH in the Cygnus X-1 binary~\cite{Yoshino:2014wwa, Sun:2019mqb, Collaviti:2024mvh} and proposed follow-up of well-localized merger remnant BHs that could be measured in LVK future observing runs, for vectors~\cite{Baryakhtar:2017ngi, Jones:2023fzz, Jones:2024fpg}, or next generation detectors, for scalars~\cite{Arvanitaki:2016qwi, Ghosh:2018gaw, Isi:2018pzk}.
In this work, we focus on vector superradiance clouds that also have an electromagnetic counterpart~\cite{Siemonsen:2022ivj} and look for the corresponding CW signals in  publicly available LVK O3 data~\cite{KAGRA:2023pio}.

Additionally, there is a third class of observational imprints arising from the dynamics of boson clouds around BHs in a binary system~\cite{Baumann:2018vus}. In such configuration, the CW signal from the cloud~\cite{Baumann:2018vus} and the GW emission from the binary~\cite{Baumann:2018vus,Baumann:2022pkl, Xie:2022uvp, Hannuksela:2018izj} can both be significantly affected, leading to potentially observable effects in next-generation GW detectors.

New light and weakly coupled vector fields constitute one of the simplest extensions of the Standard Model and are expected to arise in a variety of string theory constructions~\cite{Abel:2008ai, Arvanitaki:2009hb, Goodsell:2009xc, Camara:2011jg}. These particles, generically referred to as dark photons, could account for the dark matter abundance~\cite{Pospelov:2008jk, Arias:2012az, Nelson:2011sf, Graham:2015rva, Agrawal:2018vin, East:2022rsi} or serve as a mediator between the dark matter and the visible sector. Generically, dark photons can mix with regular photons~\cite{Okun:1982xi,Holdom:1985ag}, leading to a variety of observational effects that are targeted by a wide ongoing experimental program, including laboratory setups and ground- and space-based telescopes~\cite{Caputo:2021eaa, Antypas:2022asj}.\footnote{See~\cite{AxionLimits} for a collection of existing constraints and projections for upcoming observations in the dark photon parameter space. Notice that several bounds derived in the literature and collected in~\cite{AxionLimits} only apply if dark photons make up the totality of the dark matter abundance, while BH superradiance occurs independently of any cosmological abundance.} Dark photons with masses below $\sim 10^{-7}$ $\rm{eV}/c^2$ that do not make up a significant fraction of the total dark matter abundance are particularly elusive, as most of the observational effects in the laboratory and stellar environments are suppressed at small masses.
Precise measurements of the cosmic microwave background (CMB) spectrum~\cite{Mirizzi:2009iz, Caputo:2020bdy} and its anisotropies~\cite{McCarthy:2024ozh} are sensitive to kinetic mixing angles down to approximately $2\times 10^{-7}$ and $4\times 10^{-8}$, respectively. In the mass range of $\sim 10^{-14}-10^{-12}$ $\rm{eV}/c^2$, superradiance around stellar mass BHs offers a complementary probe of  very-weakly coupled dark photons. 

Here we perform the first search for CW signals from kinetically mixed dark photon superradiance clouds. In the presence of mixing with the Standard Model photon as small as about a few times $10^{-10}$, the cloud is expected to emit large amounts of electromagnetic (EM) radiation across a range of frequencies, potentially with a periodic component~\cite{Siemonsen:2022ivj}. The periodicity is associated with the cloud oscillation timescale, which is fixed, at leading order, by the dark photon mass $\dpm$, giving rotational periods of order $\sim \mathrm{ms}\ [10^{-12}\ \mathrm{eV}/(\dpm c^2)]$. We therefore postulate that stellar mass Galactic BHs, dressed with a superradiance cloud, could be hiding in existing catalogs as millisecond radio pulsars with nearly degenerate rotational frequencies. If this is the case, the expected CW emission from such ``new pulsars'' could be much larger than that from regular neutron stars, and a positive detection would be a smoking gun signature of dark photon superradiance. We select a set of pulsating radio sources and look for their CW counterpart, relying on the measured ephemerides; when these are well-measured, a targeted search can provide a significant enhancement in sensitivity compared to all-sky searches~\cite{KAGRA:2022dwb} [see Ref.~\cite{Tenorio:2021wmz} for more details on these searches]. We highlight that this is the first search for signals from vector superradiance clouds with nonzero nongravitational interactions and it complements the existing and upcoming searches described above that do not require an EM counterpart.

We select 44 pulsating radio sources, with rotational frequencies between 56.5 and 353.4 Hz, corresponding to GW frequencies between 113.1 and 706.7 Hz, and boson masses between $2.3\times 10^{-13}$ and $1.5\times 10^{-12}$ $\rm{eV}/c^2$. Among these, 4 sources have been targeted by a CW search before and 6 have ephemerides uncertainties that are too large to allow for an improvement in sensitivity with respect to an all-sky search.  For the remaining 34,\footnote{Note that usual LVK searches target pulsars whose ephemerides measurements overlap a given observing run. This reduces the chances of unknown stochastic processes, such as glitches~\cite{Ashton:2017wui} or accretion-induced spin wanderings~\cite{Mukherjee:2017qme}, that can lower the detection chances if not correctly handled. The boson cloud should not be affected by such processes, hence, we can loosen this condition. On top of that, searches may select targets depending on spin-down strain upper limits and computing availability (see e.g.,~\cite{LIGOScientific:2021quq,LIGOScientific:2021hvc}).} we perform the most sensitive search to date of a CW counterpart 
using a narrow-band pipeline~\cite{AstoneColla2014,Mastrogiovanni:2017xjr}, a resampling pipeline for sources in binary systems~\cite{Singhal:2019dfn} and two semicoherent pipelines~\cite{Mirasola:2024kll,DAntonio:2023jxm}. We find no statistically significant evidence for a CW signal and place 95\% confidence level upper limits (CL ULs) on the signal strain amplitude in the range $1.05\times 10^{-26}$--$ 9\times 10^{-26}$.
These bounds are, in most cases, more than an order of magnitude below the maximum dark photon cloud strain amplitude---which occurs if the hosting BH has the optimal mass and was formed around 1000 years ago---showing how a possible detection was within reach of this analysis. 

In order to interpret the strain ULs directly in terms of dark photon mass and coupling, free of assumptions on the BH mass, initial spin, and age, we estimate the expected CW signals from a population of Galactic BHs, integrating over the BH parameters. When a large number of observable events is expected, we can ``disfavor'' the corresponding dark photon mass. We find that a discrete set of masses between about $2\times 10^{-13}$ and $1.4\times 10^{-12}$ $\rm{eV}/c^2$ should have produced an observable signal in our analysis, for kinetic mixing couplings between about $10^{-9}$ and $3\times 10^{-7}$ (or as small as $3\times 10^{-10}$ for more optimistic assumptions about the maximum mass and spin of the Galactic BH population). A similar approach was used in Refs.~\cite{Arvanitaki:2014wva,Arvanitaki:2016qwi, Brito:2017zvb} and Ref.~\cite{Baryakhtar:2017ngi} to estimate the expected sensitivity of CW searches to gravitationally coupled scalar and vector superradiance signals, respectively. Reference~\cite{Zhu:2020tht} performed a detailed numerical study for scalar clouds using simulations of Galactic BHs and results from all-sky searches. The framework presented here constitutes an alternative analytic approach that can also be used to directly interpret all-sky CW searches.

The paper is organized as follows. Section~\ref{sec:dp_superrad} summarizes the physics of kinetically mixed dark photon superradiance and its observational features relevant for the current search. Section~\ref{sec:gw_search} contains the details about the CW search, with a description of the signal in Sec.~\ref{sec:signal}, the selected targets in Sec.~\ref{sec:targets}, the pipelines used in Sec.~\ref{sec:methods}, and how these are matched to each source in Sec.~\ref{subsec:targets_and_methods}; the results are presented in Sec.~\ref{sec:results}. Section~\ref{sec:interpretations} provides an interpretation of the strain ULs in terms of the dark photon parameters. Finally, we conclude in Sec.~\ref{sec:conclusion}. We also include a series of appendices containing further details and technical derivations. We use $c$ and $\hbar$ to indicate the speed of light and the reduced Planck constant, respectively.

\section{Kinetically mixed dark photon superradiance\label{sec:dp_superrad}} 

We begin by briefly introducing relevant aspects of BH superradiance, focusing specifically on ultralight vector bosons $A'_\mu$ [see Ref.~\cite{Brito:2015oca} for a review]. For the purposes of this work, the mass $\dpm$ of $A'_\mu$ is assumed to arise from a Stueckelberg mechanism~\cite{2009esuf}, and the kinetic mixing is controlled by the mixing parameter $\varepsilon$ that couples the massive vector boson to the Standard Model photon $A_\mu$, as described by the Lagrangian density
\begin{align}
\begin{aligned}
    \mathcal{L}=-\frac{1}{4}F_{\mu\nu}F^{\mu\nu} & -\frac{1}{4}F'_{\mu\nu}F'^{\mu\nu}\\
    &-\frac{1}{2}\frac{\dpm^2c^2}{\hbar^2}A'^\mu A'_\mu+\frac{\varepsilon}{2}F'_{\mu\nu}F^{\mu\nu},
\end{aligned}
\end{align}
where $F_{\mu\nu}$ and $F_{\mu\nu}'$ are the regular and dark electromagnetic field-strength tensors, respectively. Let us ignore the kinetic mixing for now and return to it below. In the purely gravitational case, the activity and efficiency of the superradiance process around a BH of mass $M$ and dimensionless spin $\chi$ is controlled by the dimensionless ``gravitational fine-structure constant''
\begin{align}
    \alpha=\frac{GM \dpm}{\hbar c},
\end{align}
where $G$ denotes Newton's gravitational constant. Through the superradiance process, a set of quasibound states of the massive vector boson around the spinning BH are populated exponentially quickly. These states are characterized by their azimuthal number $\tilde{m}$, polarization state $S$, and overtone number $\hat{n}$ [following the notation of Ref.~\cite{Dolan:2018dqv}]. Here and in the following, we focus entirely on the fastest growing state [i.e., the $(\tilde{m},\hat{n},S)=(1,0,-1)$ state]. Hence, the instability is active for all $\alpha\lesssim 1/2$ and becomes less efficient in the $\alpha\ll 1$ limit. This unstable growth saturates in the formation of a superradiance cloud with angular oscillation frequency $\omega$, surrounding the BH with horizon frequency $\Omega_H=\chi c^3[2GM(1+\sqrt{1-\chi^2})]^{-1}$, once the latter is spun down to (approximately) saturate the superradiance condition $\omega\approx\Omega_H$ \cite{Arvanitaki:2010sy, East:2017ovw,East:2018glu}. The mass of this cloud in this saturated state, $M^{s}_{c}$, may be up to $\approx 10$\% of the BH's mass-energy. 

The subsequent cloud evolution is primarily driven by GW emission. The cloud's energy is dissipated to infinity through the total GW energy flux, given by $P_{\rm GW}\sim c^5 G^{-1}\alpha^{10} M_c(\tau)^2/M^2$ in the $\alpha\ll 1$ regime \cite{Baryakhtar:2017ngi,Siemonsen:2019ebd} (see Ref.~\cite{Siemonsen:2022yyf} for accurate numerical coefficients of these expressions and higher-order corrections), where $M_c(\tau)$ is the time-dependent cloud mass.\footnote{Here and in the following, we will use $\tau$ to denote time in the cloud reference frame and $t$ for time in the detector frame.} This process takes place over a GW timescale $\tau_{\rm GW}$ and leads to a polynomial-type decay of the cloud's mass:
\begin{align}
    M_{c} (\tau) = \frac{M^{s}_c}{1+\tau/\tau_{\rm GW}},
    \label{eq:cloudmassevo}
\end{align}
where $\tau=0$ corresponds to superradiance instability saturation, such that $M_c(0)=M^s_c$. Explicitly, the GW emission timescales are approximately \cite{Siemonsen:2019ebd}
\begin{align}
    \tau_{\rm GW} \approx 9\times 10^9 \ \textrm{years}\left(\frac{M}{10M_\odot}\right)\left(\frac{0.01}{\alpha}\right)^{11}\frac{1}{\chi},
    \label{eq:tgw}
\end{align}
where $\chi$ is the dimensionless BH spin.

For a BH at a distance $d$ and following the conventions of \texttt{SuperRad} \cite{Siemonsen:2022yyf}, we can define a characteristic strain of the emitted GW measured at the detector as $h_0(\tau) = \sqrt{10 G/c^3 P_{\rm{GW}}(\tau)}/(2\pi f_{\rm GW} d)$~\cite{Siemonsen:2022yyf}, which also decays in time from the maximum value at saturation, $h^s_0$, according to Eq.~\eqref{eq:cloudmassevo}:
\begin{equation}
    h_0 (\tau) = \frac{h^s_0}{1+\tau/\tau_{\rm GW}}.
    \label{eq:strainevo}
\end{equation}
The dominant contribution to the gravitational waveform is the quadrupole. For sufficiently large $\alpha$, higher-order modes become important [see Ref.~\cite{East:2017mrj,Siemonsen:2019ebd}]. However, this parameter regime is irrelevant for this work, as our signal is dominated by relatively long-lived (small-$\alpha$) systems. The waves emitted are quasi-monochromatic with frequency $f_{\rm GW}=\omega/\pi$, where (for the fastest-growing state) the cloud's angular oscillation frequency is \cite{Baryakhtar:2017ngi,Siemonsen:2022yyf}
\begin{align}
    \omega = \frac{\dpm c^2}{\hbar}\left\lbrace 1 - \left[\frac{\alpha^2}{2} + \mathcal{O}\left( \alpha^4 \right)\right] - \left[\frac{5\alpha^2}{8} + \mathcal{O}\left( \alpha^3 \right)\right]\frac{M_{c}}{\M} \right\rbrace .
    \label{eq:omega}
\end{align}
To zeroth order, the GW frequency is given by the mass of the ultralight vector boson $\dpm$. The first correction in Eq.~\eqref{eq:omega} is due to the gravitational potential of the BH, while the third term is due to the gravitational potential of the superradiance cloud itself. The latter is time-dependent according to Eq.~\eqref{eq:cloudmassevo}, and therefore, the dissipation of the cloud's mass results in a mild increase of the GW frequency over time~\cite{Arvanitaki:2014wva} [see also Ref.~\cite{May:2024npn} for an accurate treatment of this frequency shift]. At the point of saturation, the first derivative of the GW frequency is roughly given by \cite{Siemonsen:2022yyf}
\begin{align} \label{eq:fdotGW_0}
    \dot{f}^s_{\rm GW}\approx 10^{-21} \ \textrm{Hz/s}\left(\frac{10 M_\odot}{M}\right)^2\left(\frac{\alpha}{0.01}\right)^{15}\chi^2
\end{align}
in the $\alpha\ll 1$ limit. This spin-up rate decreases as the gravitational self-energy of the cloud reduces due to the GW emission, as $\dot{f}_{\rm GW} \propto \dot{M}_c(\tau)$.

We return now to the impact of a nonvanishing kinetic mixing angle $\varepsilon$. This was studied in detail in Ref.~\cite{Siemonsen:2022ivj}.\footnote{Here, as in Ref.~\cite{Siemonsen:2022ivj}, we focus on non-accreting BHs. See also the recent work from Ref.~\cite{Xin:2024trp}, where initial steps towards understanding the impact of accretion disks were considered. To the best of our knowledge, there is no evidence for accretion for any of the pulsar sources targeted in our search.} The massive vector field $A'_\mu$ sources visible electromagnetic fields that act on charged particles around the BH. Early on in the exponential growth of the superradiance process, these fields are sufficiently strong to produce electron-positron pairs through the photon-assisted Schwinger mechanism~\cite{Dunne:2009gi,Monin:2010qj}, leading to a pair production cascade and, ultimately, the formation of a tenuous pair plasma within the cloud around the BH. This plasma enters a highly turbulent, but nearly force-free, state with the ambient electromagnetic fields. In this state, large amounts of electromagnetic energy are injected into the plasma through dissipative processes (such as at magnetic reconnection sites) in the bulk of the cloud. This energy injection drives the electromagnetic power output of the system with a total luminosity of~\cite{Siemonsen:2022ivj}
\begin{equation}\label{eq:lum}
    L \approx 0.13 \varepsilon^2 \alpha \frac{c^5 M_c}{G\M}
\end{equation}
in the $\alpha\ll 1$ regime. For $\varepsilon\sim10^{-7}$, corresponding to the largest observationally allowed coupling in the mass range considered, this luminosity may reach $10^{42}$ erg/s. Note also, this electromagnetic power output decreases only on long timescales dictated by Eq.~\eqref{eq:tgw} since $L\propto M_c(\tau)$. For sufficiently large $\varepsilon$ or small $\alpha$, the cloud's evolution after saturation is no longer driven primarily by GW emission, but is affected also by the large electromagnetic dissipation. However, in what follows, we restrict to the dark photon parameter space where the subsequent evolution is dictated by the gravitational radiation alone and the cloud decays according to the power law given in Eqs.~\eqref{eq:cloudmassevo} and \eqref{eq:strainevo}. 

A significant fraction of this power output is likely emitted in the form of x-/$\gamma$-rays, while a small fraction may also leave the system in the form of radio waves. Due to the turbulence of the plasma and the nature of the resistive processes, much of the electromagnetic power is emitted isotropically. However, some evidence for the presence of a periodic component in the emission was found in Ref.~\cite{Siemonsen:2022ivj}. The frequency of the periodicity of this component is set by the dark photon's mass and is half of the GW frequency: $f_{\rm EM}=\omega/(2\pi)$. As a result, for stellar mass BHs, these kinetically mixed superradiance clouds may disguise themselves as ordinary millisecond pulsars. The fraction of radio emission and the beaming factor cannot be established precisely within the formalism used in Ref.~\cite{Siemonsen:2022ivj}.\footnote{What can be inferred about the cloud emission spectra and periodicity is further discussed in Sec. IV of Ref.~\cite{Siemonsen:2022ivj}.}
In analogy with the pulsar, we assume that a fraction $f_r$ of the electromagnetic power output is emitted in the form of radio waves modulated by the periodicity set by the dark photon's mass. 
We believe this analogy is sufficient to motivate a search for CWs from pulsating radio sources that could be associated with dark photon clouds. This is further encouraged by the fact that analysis pipelines for these types of searches have already been developed and are routinely applied to LVK data [see e.g.~\cite{LIGOScientific:2021quq, 2017PhRvD..96l2006A, O2Narrowband, 2015PhRvD..91b2004A,Singhal:2019dfn,DAntonio:2023jxm,Mirasola:2024kll}] to look for CWs emitted by regular neutron stars. Applying a similar search to a broader set of targets could potentially reveal a dark photon signal. In the next section, we describe the expected signal and the selected targets in more detail, together with the results of the search.

A series of projections and  searches for superradiant vector clouds in the gravitational regime have been undertaken in recent years, as outlined in the Introduction. The analyses to date have relied purely on the gravitational dynamics and signatures of the cloud. Since the superradiance process is impacted by the kinetic mixing perturbatively in the $\varepsilon\ll 1$ regime, these search results are effectively assuming a  vanishing kinetic mixing of the underlying superradiating dark photon; i.e., they are only valid below a critical mixing strength, $0\leq\varepsilon <\varepsilon_c$. The critical $\varepsilon_c$ depends on the type of search performed and target considered. For instance, in the case of binary BH follow-up searches at future GW observatories,  Ref.~\cite{Jones:2024fpg} determined $\varepsilon_c\sim 10^{-6}$--$10^{-4}$ depending on the target.  Importantly, BH spin measurements based on electromagnetic observations cannot be robustly interpreted at nonzero $\varepsilon$, as the impact of a nonvanishing kinetic mixing on the accretion flow, electromagnetic emission, and evolution of the superradiance process---relevant for BH measurements in x-ray binaries---is yet to be comprehensively understood [initial steps were taken in Ref.~\cite{Xin:2024trp}]. The results obtained in this work, on the other hand, require a nonvanishing kinetic mixing of $\gtrsim\mathcal{O}(10^{-10})$ for the efficient creation of charged particles in the cloud, making this approach complementary to existing searches for superradiant vector clouds.

\section{Gravitational wave search} \label{sec:gw_search}

CWs are long-lasting GWs that are quasi-monochromatic in the source frame. The search method used to look for CWs emitted by the traditional observational target of rotating masses with a time-varying quadrupolar moment~\cite{Wette:2023dom}, can also be applied to look for signals emitted by boson clouds around spinning BHs.

In this section, we describe the CW search performed and the results obtained. A brief introduction to the expected signal at the detector is given in Sec.~\ref{sec:signal}, including a description of how to properly account for the relative motion between the source and the observer. Section~\ref{sec:targets} outlines the criteria used to select radio-pulsing sources from the Australia Telescope National Facility (ATNF) pulsar catalog~\cite{Manchester:2004bp,pulsarlist}, which may be associated with dark photon superradiance (see Sec.~\ref{sec:dp_superrad}). Next, Sec.~\ref{sec:methods} provides an overview of the methods employed to search for a CW counterpart. The characteristics of each target (e.g. accuracy of the measured ephemerides, being isolated or in a binary system) are accounted for by selecting the appropriate method as detailed in Sec.~\ref{subsec:targets_and_methods}. Our results are presented in Sec.~\ref{sec:results}. The interpretation of these results is provided in the next section, Sec.~\ref{sec:interpretations}.

\subsection{Signal at the detector\label{sec:signal}}
The persistence of CWs can be exploited to build up the signal-to-noise ratio (SNR) by analyzing long stretches of data. 
The analysis of year-long datasets makes it mandatory to account for the relative motion of the detector with respect to the source.
Following Ref.~\cite{Jaranowski:1999pd}, the detector's response to a quadrupolar CW can be expressed as 
\begin{align}
    \label{Eq:hSignal}
    h(t) = & h_0(\tau_{\rm ref})\,F^+(t;\vec{n},\Psi)\,  \frac{1+\cos^2\iota}{2}\, \cos \phi(t) +\nonumber\\
    & h_0(\tau_{\rm ref})\,F^\times(t;\vec{n},\Psi)\, \cos\iota\,\sin \phi(t), 
\end{align}
where $\iota$ is the inclination angle of the BH rotational axis with respect to the line of sight and $F^{+, \times}$ are the time- and detector-dependent beam pattern functions that describe the detector response as a function of the source location $\Vec{n}$ with respect to the Solar System barycenter (SSB) and polarization angle $\Psi$. 
The GW phase $\phi(t)$ includes the Doppler modulation linked to the relative motion of source and detector.
The CW amplitude $h_0(\tau_{\rm ref})$ is the characteristic strain of the cloud emission at the reference time $\tau_{\rm ref}$ when the observation starts. The strain amplitude depends on the dark photon mass and BH properties (mass, initial spin, age, and distance), and is given in Eq.~\eqref{eq:strainevo}.

In our analyses, we use a slightly different formalism [see Ref.~\cite{Astone:2010zz}] to describe the CW at the detector, in which Eq.~\eqref{Eq:hSignal} is replaced by an equivalent expression given by the real part of
\begin{equation}
\label{eq:hSignal_HplusHcross}
    h(t)  = H_0\, (A^+ H_+ + A^\times H_\times)\, e^{j\phi (t)},
\end{equation}
where $A^{+/\times}(t;\Vec{n}) = F^{+/\times}(t;\Vec{n}, \Psi=0)$, and
\begin{align}
    H_0 & = \frac{h_0(\tau_{\rm ref})}{2} \sqrt{1 + 6\, \cos^2\,\iota + \cos^4\,\iota}\,, \label{eq:Capital_H}\\
    H_{+} & = \frac{\cos\,2\Psi\>- j\, \eta \,\sin\,2\Psi}{\sqrt{1+\eta^2}}\,,\label{eq:Hp}\\
    H_{\times} & = \frac{\sin\,2\Psi\>+ j\, \eta \,\cos\,2\Psi}{\sqrt{1+\eta^2}}\,,\label{eq:Hc}
\end{align}
where $\eta$ is the ratio of the semiminor to the semimajor axis of the polarization ellipse,
\begin{equation}
     \eta = -\frac{2\, \cos\,\iota}{1+\cos^2\,\iota}.
\end{equation}
Moreover, since the frequency of the signal varies slowly over time (see the discussion in Sec.~\ref{sec:dp_superrad} and later in Sec.~\ref{sec:targets}), we can expand the phase evolution in the source reference frame as
\begin{equation}
        \label{Eq:dopp_mod}
        \frac{\phi^{\rm src}(\tau)}{2\pi} = \frac{\phi_0}{2\pi} + f_{\rm GW} (\tau - \tau_{\rm ref}) + \frac{\Dot{f}_{\rm GW}}{2} (\tau - \tau_{\rm ref})^2 + ... ,        
\end{equation}
where $\tau$ is the time at the source, $\phi_0$ is the initial phase of the signal, and $\Dot{f}_{\rm GW}$ is the frequency derivative at $\tau_{\rm ref}$. Since source and detector have a relative motion, the induced Doppler effect can be described as a delay between $\tau$ and the arrival time at the detector, $t_{\rm arr}$.
In this sense, the Doppler modulation removal consists of relating $\tau$ to $t_{\rm arr}$ such that $\phi(t_{\rm arr})= \phi^{\rm src}(\tau(t_{\rm arr}))$. Following Eq.~(5) of~\cite{Leaci:2016oja}, $\tau(t_{\rm arr})$ can be written for an isolated emitter as
    \begin{align}
        \label{eq:source_proper_time}
        \tau(t_{\rm arr}) = t_{\rm arr} + \frac{\Vec{r}_{\rm SSB}\cdot \Vec{n}}{c} -  \Delta_E - \Delta_S\, - \frac{d}{c},
    \end{align}
where $\Vec{r}_{\rm SSB}$ is the Earth's position in the SSB at time $t_{\rm arr}$; thus, the second term represents the vacuum delay between the arrival time at the observatory and the SSB, usually referred to as R\o{}mer delay. The redshift due to the other Solar System bodies is taken into account by the Einstein delay ($\Delta_E$). The term $\Delta_S$ is the Shapiro delay, linked to the deflection of a signal passing close to the Sun. The last term is the signal (light) vacuum travel time from the source, at a distance $d$, to the SSB. Following~\cite{Wette:2023dom,Jaranowski:1999pd}, $d/c$ can be reabsorbed in $\tau_{\rm ref}$. 

If the source is located in a binary system, the orbital motion induces an additional time-dependent delay to Eq.~\eqref{eq:source_proper_time} as
\begin{align}
    \label{eq:source_proper_time_bin}
    \tau_{\rm bin} (t_{\rm arr}) = \tau(t_{\rm arr}) - \frac{R(\tau_{\rm bin})}{c} - \Delta_E^{\rm bin} - \Delta_S^{\rm bin}\,,
\end{align}
where $\tau_{\rm bin}$ is the time at the source in this scenario, $\Delta_E^{\rm bin}$ and $\Delta_S^{\rm bin}$ denote the Einstein and Shapiro delays due to the binary companion and possible planets in the system, respectively~\cite{Tempo2_1,Tempo2_2}. Lastly, $R(\tau_{\rm bin})/c$ is the binary R\o{}mer delay with $R(\tau_{\rm bin})$ being the radial distance of the source from the binary barycenter (BB), projected along the line of sight [we refer the reader to Refs.~\cite{Leaci:2016oja,Tempo2_2} for its expression].

Due to Earth's rotation, the time dependence of the detector response to a quasi-monochromatic GW splits it into five harmonics~\cite{Astone:2010zz} at $f_{\rm GW}\,,f_{\rm GW}\pm f_\oplus\,,f_{\rm GW}\pm 2 f_\oplus$, where $f_\oplus\simeq 1.16\times 10^{-5}$~Hz is the Earth sidereal frequency. When the frequency resolution of the search methods approaches $f_\oplus$, this effect must be taken into account, as we do in the searches outlined below. Other searches based on short-segment integration can neglect it without loss of sensitivity~\cite{Mirasola:2024kll,Astone:2014esa}.

\subsection{Target selection\label{sec:targets}}

We select a subset of radio pulsating sources from the ATNF pulsar catalog~\cite{Manchester:2004bp,pulsarlist} to search for a CW signal, based on the criteria identified in Ref.~\cite{Siemonsen:2022ivj}.
The superradiance cloud of a kinetically-mixed dark photon exhibits two characteristic features that are crucial for defining our targets. First, the rotational period is tied to the dark photon mass [see Eq.~\eqref{eq:omega}]. Second, the decay of the cloud leads to a spin-up of the system over time [see Eq.~\eqref{eq:fdotGW_0}].
The ``anomalous'' pulsating sources would therefore be multiple sources with the same intrinsic rotational frequency and sources with a positive intrinsic frequency derivative. The quantities observed by a radio telescope are\footnote{Note that the radio observations do not measure directly the intrinsic quantities, due to the relative motion between the source and the observer.}
\begin{align}
    \label{eq:obs_freq}
    \left. f_{\rm EM}\right|_{\rm obs} & = \frac{\left. f_{\rm GW}\right|_{\rm obs}}{2} = \frac{\dpm c^2}{2\pi \hbar}\left(1-\frac{v_{\rm rad}}{c} \right)  \\
    \label{eq:obs_fdot}
    \left.\frac{\dot{f}_{\rm EM}}{f_{\rm EM}}\right|_{\rm obs} & = \left.\frac{\dot{f}_{\rm GW}}{f_{\rm GW}}\right|_{\rm obs} \simeq - \frac{5\alpha^2}{8}\frac{\dot{M}_{c}}{\M} - \frac{a_{\rm rad}}{c} - \frac{\mu_{\rm pm}^2 d}{c}, 
\end{align}
where $\dpm$ is the boson mass [and we have taken the leading order approximation of Eq.~\eqref{eq:omega}, $\omega \simeq \dpm c^2/\hbar$, since the bulk of the signals is dominated by long-lived, small-$\alpha$ systems, see Sec.~\ref{sec:CWh_dist}], $v_{\rm rad}$ and $a_{\rm rad}$ are the source's line-of-sight velocity and acceleration as measured by the observer (which give a Doppler shift correction), while $\mu_{\rm pm}$ denotes the source's proper motion and the corresponding term in Eq.~\eqref{eq:obs_fdot} is known as the Shklovskii effect. For BHs in the Galaxy, the typical velocity will be of order $v_{\rm rad}/c \sim \mathcal{O}(10^{-3})$.\footnote{In the simulations of galactic BHs in~\cite{Zhu:2020tht}, $90\%$ of BHs had $|v_{\rm rad}/c| < 6\times 10^{-4}$ and the maximum value was found to be $|v_{\rm rad}/c| = 0.0025$. The additional corrections to the frequency due to the gravitational potential of the BH and the cloud itself (from Eq.~\eqref{eq:omega}) are smaller than the typical Doppler shift.} 

We therefore select two classes of ``anomalous'' pulsars based on the two features described above. The first class is frequency multiplets (doublets or triplets), defined as the sources $i$ with at least one other pulsar $j$ that satisfies $|f_i -f_j|/f_i < v_{\rm rad}/c \simeq 10^{-3}$, to account for Doppler shifts. The second group is sources that are apparently spinning up, {\it i.e.} with $\left. \dot{f}_{ \rm EM}\right|_{\rm obs} > 0$. In this last category, we exclude pulsars that are known to be in a binary system, to avoid spurious positive spin frequency derivatives from large accelerations. The measured spin-up rates are within the range $[4.2 \times 10^{-17}, 1.1 \times 10^{-14}]$ Hz/s, which can be obtained for BH masses below $20\ M_{\odot}$ (although the spin-up rate observed today also depends on the age of the BH).

\begin{figure}[!ht]
    \centering 
    \includegraphics[width=0.5\textwidth]{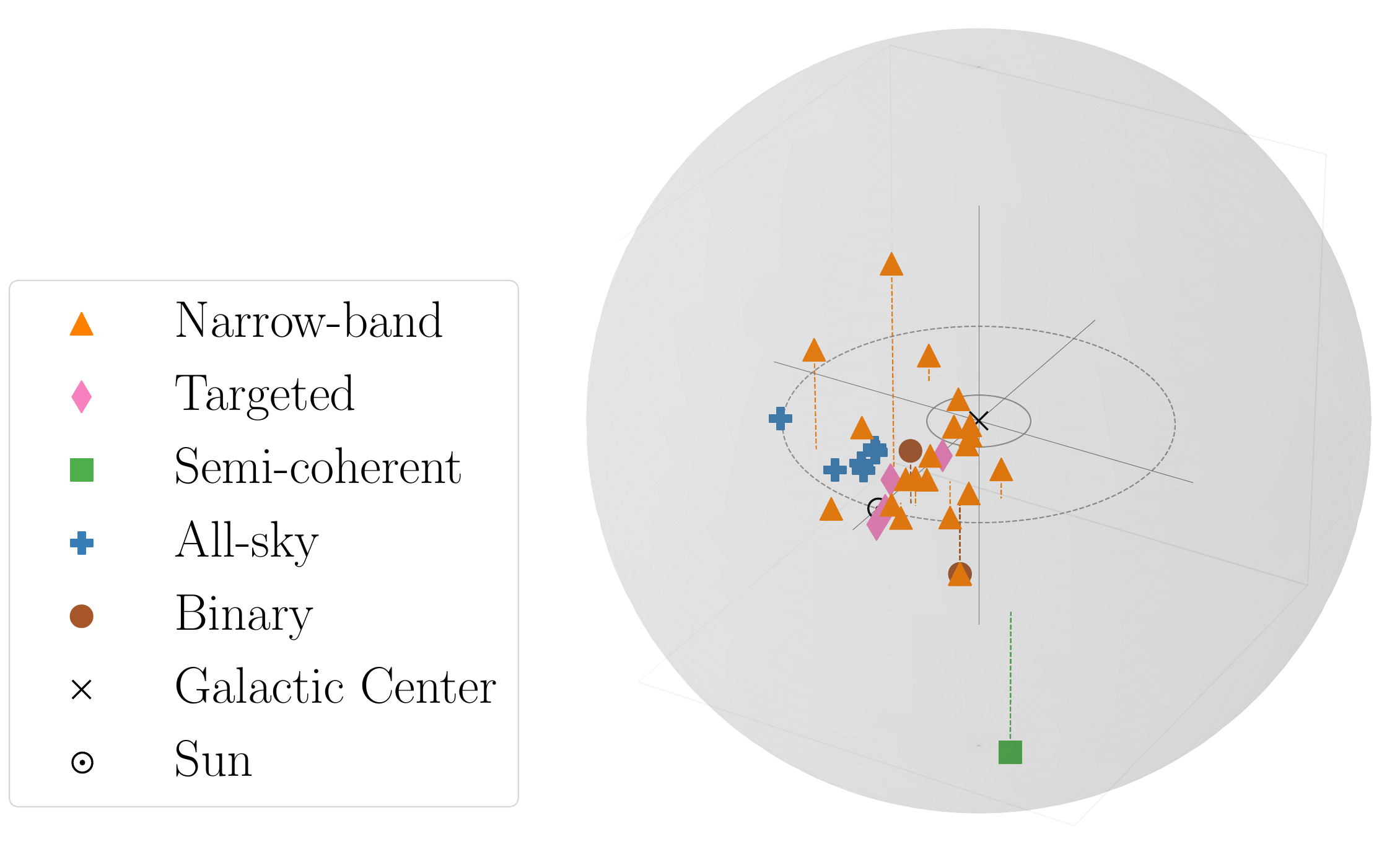}
    \caption{Three-dimensional representation of the 44 selected radio sources, shown in galactocentric coordinates (see Table~\ref{tab:results}: note that some sources are close to each other and the markers overlap). For each source, the color and shape of the marker represent the analysis pipeline used, as described in Sec.~\ref{sec:methods}. Among these sources, 34 are studied for the first time in this work using the narrow-band pipeline (orange triangles), the resampling pipelines for sources in a binary system (brown circles), and the semicoherent pipeline (green squares). Additionally, for 4 sources we use results from the targeted search of Ref.~\cite{LIGOScientific:2021hvc} (pink diamonds) and for 6 others results from the all-sky search of Ref.~\cite{KAGRA:2022dwb} (blue crosses).
    The vertical lines correspond to the distance of the source from the Galactic plane. The $\times$ and $\odot$ symbols denote the location of the Galactic Center and the Sun, respectively. The dotted line corresponds to the approximate orbit of the Sun in the Galaxy at $\simeq 8$ kpc, while the solid line denotes the characteristic scale of the Galactic disk ($r_d = 2.15$ kpc, see Sec.~\ref{sec:galactic_bh}). 
    The gray shaded sphere has a radius of $\sim 16$ kpc, and represents the largest distance of any source from the Galactic Center.}\label{fig:sources_map}
\end{figure}

Using the above criteria, we select 59 sources from the ATNF pulsar catalog ~\cite{pulsarlist, Manchester:2004bp}. Of these, we discard 15 sources in binary systems that do not have accurate enough ephemerides to produce reliable results. The remaining 44 sources' positions within the Galaxy are shown in Fig.~\ref{fig:sources_map} and labeled with the corresponding search used (see Secs.~\ref{sec:methods} and~\ref{subsec:targets_and_methods}), while the full properties for each source are listed in Tabs.~\ref{tab:pos_fdot},~\ref{tab:doublets}, and~\ref{tab:triplets} in Appendix~\ref{app:source_params}. Among these sources, 4 have already been targeted by previous searches~\cite{LIGOScientific:2021hvc} and their results will be used in this work. Additionally, 6 isolated sources are not measured well enough for a search sensitivity better than the all-sky search reported in Ref.~\cite{KAGRA:2022dwb}. We therefore analyze for the first time the remaining 34 sources, using the analysis pipelines outlined in the next section. We use the ephemerides provided by the ATNF catalog, ensuring that the timing solutions are the most up-to-date in the literature, while for two pulsars (J1122-3546 and J1317-0157), we use the ephemerides reported in Ref.~\cite{Swiggum:2022xlb}. To summarize, in this work, we use 44 pulsating sources to perform a kinetically mixed dark photon search, 34 of which are analyzed in CWs for the first time.

\subsection{Methods\label{sec:methods}}

Due to uncertainties in the source parameters, a tailored pipeline is required for the CW search. We account for both sky location uncertainties and, when needed, binary orbital ones, as discussed in Ref.~\cite{Mirasola:2024kll}.\footnote{Uncertainties in rotational parameters are negligible for the targets considered.} Searches are typically divided into fully coherent and semicoherent types. The former analyzes data in a single segment (typically with timespans of $T_{\rm obs}\approx$~1~year) and is more sensitive~\cite{Astone:2014esa,Wette:2023dom}. In semicoherent searches, data are split into shorter segments (coherence time, $\Tfft$) and combined incoherently. While this reduces sensitivity, it enhances robustness against uncertainties.

Below, we briefly outline the methods used in our search. We employ two fully coherent pipelines when sufficiently accurate sky localization is available.
The narrow-band pipeline, detailed in Sec.~\ref{subsec:narrowband}, is used for isolated targets. For sources in binary systems, we apply the resampling method described in Sec.~\ref{sec:resamp_pipeline}, which effectively accounts for uncertainties in orbital parameters. Finally, for isolated targets with large position uncertainties, we use the semicoherent method outlined in Sec.~\ref{subsec:5vectsemi}. The criteria for selecting the appropriate method for each target are discussed in Sec.~\ref{subsec:targets_and_methods}.

\subsubsection{Narrow-band pipeline\label{subsec:narrowband}}

The narrow-band pipeline described in~\cite{AstoneColla2014,Mastrogiovanni:2017xjr}, and used in several searches [see e.g.~\cite{LIGOScientific:2021quq, 2017PhRvD..96l2006A, O2Narrowband, 2015PhRvD..91b2004A}] looks for a CW emitted by an isolated source at $f_{\rm GW} (t) \approx 2\,f_{\rm EM} (t)$ by applying waveform templates that span a grid of values in the $f_{\rm GW}-\Dot{f}_{\rm GW}$ parameter space. The range covered is typically of the order of a few tenths of Hz in frequency and $\mathcal{O}(10^{-14})$~Hz/s in the frequency derivative. The sensitivity of the method depends on the number of templates explored~\cite{AstoneColla2014}, which is usually of order $\mathcal{O}(10^{6}-10^8)$.

Starting from the data in the time domain~\cite{Astone:2005fj}, for each target we perform a Doppler correction with nonuniform resampling that is independent of the CW frequency with successive downsampling at 1~Hz.  Then, we look for the sidereal modulation imprinted by Earth's rotation, which splits the signal into five harmonics $f_{\rm GW}\,,f_{\rm GW}\pm f_\oplus\,,f_{\rm GW}\pm 2 f_\oplus$; see the discussion in Sec.~\ref{sec:signal}. Using two matched filters (for the CW polarizations), we construct the detection statistic
\begin{equation}
    \mathcal{S}=|\hat{H}_+|^2|A^+|^4+|\hat{H}_\times|^2|A^\times|^4,
    \label{eq:s_statistic}
\end{equation}
where $\hat{H}_{+/\times}$ are the GW amplitude estimators~\cite{Mastrogiovanni:2017xjr}
\begin{equation}
    \hat{H}_{+/\times} = \frac{\textbf{X}\cdot \textbf{A}^{+/\times}}{|\textbf{A}^{+/\times}|^2}
\end{equation}
and where the bold notation identifies the so-called \textit{5-vector}, consisting of the Fourier components of the data $X$ and the sidereal responses $A^{+/\times}$ at the five frequencies of interest. Following~\cite{Mastrogiovanni:2017xjr}, we coherently combine the matched filter results from all the detectors involved in the analysis. 

Lastly, from the $\mathcal{S}$ distribution, we select local maxima every 10$^{-4}$~Hz and over the considered $\dot{f}_{\rm GW}$ range. All values above a certain threshold $\mathcal{S}_{\rm thr}$, set by fixing the false alarm probability (FAP) at 1\% taking into account the number of trials, are followed up. The noise-only distribution, used to set $\mathcal{S}_{\rm thr}$, is directly inferred with an exponential fit from the tail of the histogram of all statistic values which have not been selected as local maxima. If no CW-related outlier is found, we calculate the 95\% CL ULs $h_{\rm UL}$ by injecting simulated signals in the real data.

\subsubsection{Resampling pipeline for sources in binary systems\label{sec:resamp_pipeline}} 

Targets in binary systems are more challenging to study due to the additional Doppler modulation introduced by the orbital motion of the companion star. If the orbital parameters are not known with sufficient precision, a fully coherent single-template search may suffer a significant loss in the reconstructed SNR~\cite{Leaci:2015bka,Leaci:2016oja, Mirasola:2024kll}. To address this issue, we adopt the method outlined in~\cite{Singhal:2019dfn}, where the \textit{5-vector} technique is adapted to account for uncertainties in the orbital parameters. Consequently, we perform a search in a region of parameter space surrounding the measured orbital parameters, as detailed in Appendix~\ref{app:resamp_grid}.

For every target, parameter-space point, and detector data, we apply nonuniform resampling with successive downsampling at 1 Hz (as described in Sec.~\ref{subsec:narrowband}).

Once the time series is demodulated, we construct the detection statistic $CR$ (critical ratio)\footnote{This statistic is referred to as $\mathcal{S}'$ in~\cite{Singhal:2019dfn}. Here, we adopt the $CR$ convention to standardize the notation across the pipelines.} as in~\cite{Singhal:2019dfn}, which is given by the following expression:
\begin{equation}
CR = \frac{\mathcal{S}-\mu}{\sigma}\,,
\label{eq:CR}
\end{equation}
where $\mathcal{S}$ is defined in Eq.~\eqref{eq:s_statistic} and $\mu$ and $\sigma$ are the mean and standard deviation of $\mathcal{S}$ across a given 1 Hz band, respectively.

From the detection statistic, we select candidates using a two-step veto procedure. The first veto applies a threshold on the detection statistic at $CR=18$, which corresponds to a normalized FAP of 1\% in Gaussian noise, see Ref.~\cite{Singhal:2019dfn} for details. For the candidates that survive the first veto, a double-step coincidence veto is applied. For each candidate, we require at least 3 peaks above the threshold at frequencies $f_{\rm GW}+mf_{\oplus},\;m\in[-4,4]$\footnote{The cross-correlation between the five-peaks template and the signal results in a nine-peaked detection statistic~\cite{Singhal:2019dfn}.} in every detector. To be conservative, we introduce a tolerance by considering an additional frequency bin $\Delta f= 1/T_{\textrm{obs}}$ on either side of the candidate frequency. Lastly, if no detection can be claimed, ULs at the 95\% confidence level are set through injections in real data.

\subsubsection{Semicoherent pipeline for isolated targets\label{subsec:5vectsemi}}

When parameters of isolated sources are not known with sufficient accuracy, fully coherent methods can become computationally infeasible. In such cases, an alternative is to use semicoherent pipelines, in which the whole dataset is divided into segments of shorter duration, which are individually processed and then combined without imposing signal phase coherence. 

The semicoherent method we use in this paper is still based on the concept of the \textit{5-vector}, introduced in Sec.~\ref{subsec:narrowband}, and is described in detail in Ref.~\cite{DAntonio:2023jxm}. For each given target, the data over a 0.1 Hz band around its frequency are considered separately for each detector. First, a \textit{rough}, but computationally efficient, correction of the Doppler and spin-up effects is performed, properly heterodyning the selected data over the whole observation time at coarse values of frequency and spin-up. Then, data are divided into segments of duration equal to an integer number of sidereal days (20 sidereal days in this specific case, to keep the maximum possible frequency error due to the source position uncertainty within the frequency bin) and for each of these, and for each frequency, the \textit{5-vector} statistic defined in Eq.~\eqref{eq:s_statistic} is computed. A time-frequency map of the statistic is then produced and, on this, a refined correction of the residual Doppler and spin-up effects is done, by properly shifting the frequency of the time-frequency pixels [see Ref.~\cite{DAntonio:2023jxm} for more details].

As a final step, the statistics are summed along the time axis of the corrected time-frequency map, obtaining a total statistic $\mathcal{S}(f)$ as a function of the frequency. By considering the mean $\mu$ and standard deviation $\sigma$ of the total statistic, we make use of a critical ratio, defined as in Eq.~\eqref{eq:CR}, to select outliers. Coincidences among outliers found in the detectors involved in the analysis are done, based on their a-dimensional distance~\cite{DAntonio:2023jxm}
\begin{equation}
    D = \sqrt{\left(\frac{\Delta f }{\delta f \times \Delta}\right)^2+\left(\frac{\Delta \dot f}{\delta \dot f}\right)^2}
\end{equation}
where $\Delta f,~\Delta \dot f$ are the absolute difference of the two outliers' parameters, $\delta f,~\delta \dot f$ are the resolutions, and the factor $\Delta$ weights in the proper way the distance that can be as large as four sidereal frequency bins. See~\cite{DAntonio:2023jxm} for more details. 
Any coincident outlier is subject to a follow-up step, in which the same analysis procedure is applied in a small volume around its parameters with an increased data segment duration. If no outlier survives, a 95\% CL UL on the strain amplitude is computed by injecting simulated signals in the data.

\subsection{Matching targets and methods\label{subsec:targets_and_methods}}
When the source parameters are not well-determined, a nonoptimized method may experience a significant loss in SNR, potentially missing a true detection. To address this, we apply the three methods described above, taking into account the uncertainties in sky location and binary parameters where necessary. A summary of the number of sources analyzed with each method is shown in Table~\ref{tab:methods} and will be explained in this section.

\begin{table}[ht!]
    \centering
    \renewcommand{\arraystretch}{1.3}
    \caption{Number of sources analyzed with each method discussed in Sec.~\ref{sec:methods} and the corresponding reference. Starred values indicate the results taken from previous searches. See the text for more details.}
    \label{tab:methods}
    \begin{tabular}{ccc}
        \hline \hline \rule{0pt}{3ex}  
        Method & Sources & Ref. \\
        \hline
        Narrow-band & 29 & \cite{AstoneColla2014,Mastrogiovanni:2017xjr}\\
        \hline
        Binary & 3 & \cite{Singhal:2019dfn,Mirasola:2024kll} \\
        \hline
        Semicoherent & 2 & \cite{DAntonio:2023jxm} \\
        \hline
        Targeted$^*$ & 4 & \cite{LIGOScientific:2021hvc}\\
        \hline
        All-sky$^*$ & 6 & \cite{KAGRA:2022dwb} \\
        \hline\hline
    \end{tabular}
\end{table}

Regarding the uncertainties in the sky location, we follow the approach outlined in Appendix~D of~\cite{Mirasola:2024kll}. We compute the maximum segment length, $\Tfft^{\rm sky}$, below which the SNR loss is negligible. If $\Tfft^{\rm sky} \gg T_{\rm obs}$, a single full-coherent search can be applied without needing to account for the uncertainties in the source position. Remarkably, only 8 out of the 37 isolated targets that need to be analyzed have $\Tfft^{\rm sky}<T_{\rm obs}$. Of these 8 sources, 6 have $\Tfft^{\rm sky}\ll T_{\rm obs}$, with values comparable to those used in the O3 \textit{Frequency-Hough} (FH) \textit{all-sky} search\footnote{Searches with short coherence times that explore a vast parameter space for CWs.} [see~\cite{KAGRA:2022dwb} and references therein]. For these sources, we derive the strain ULs interpolating the all-sky UL curve $h_{\rm sens}^{\rm FH}$ from~\cite{KAGRA:2022dwb} at the target frequencies. The remaining 2 sources have $\Tfft^{\rm sky}\sim51$~days and 132 days, respectively. For these, we apply the method outlined in Sec.~\ref{subsec:5vectsemi}, which offers better sensitivity than the all-sky search~\cite{Astone:2014esa}, due to its longer coherence time of 20 days versus $\mathcal{O}$(10$^3$~s) seconds from~\cite{KAGRA:2022dwb}.\footnote{The sensitivity of a semicoherent method scales as $\Tfft^{-1/4}$.} For all other isolated targets, we apply the narrow-band method described in Sec.~\ref{subsec:narrowband}.

\begin{figure*}[!hpt]
    \centering
    \includegraphics[width=0.8\textwidth]{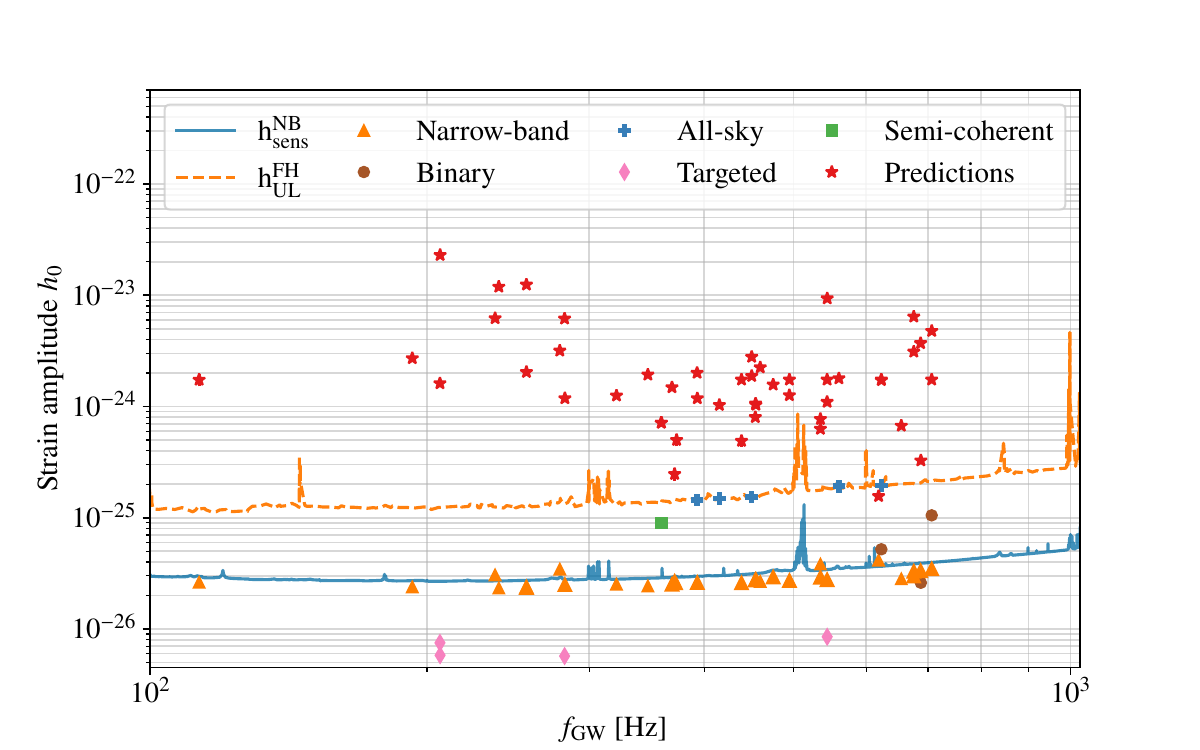}
    \caption{ULs at the 95\% CL given for all the sources presented in Sec.~\ref{sec:targets} obtained in this work using the narrow-band pipeline (orange triangles), resampling pipeline for sources in a binary system (brown circles), and the semicoherent pipeline (green squares) on O3 data from H1 and L1 detectors. We additionally show upper bounds for sources that have been targeted before [pink diamonds, \cite{LIGOScientific:2021hvc}] and upper bounds obtained interpolating the limit from the O3 all-sky search [blue crosses, \cite{KAGRA:2022dwb}]. Notice that some sources have very similar frequencies and the markers might overlap (see Table~\ref{tab:results} for the numerical results). 
    The reported $h_0^{95\%}$ ULs are smaller than the largest possible signal predicted for a dark photon cloud, assuming it is as young as $10^3$ yr [red stars, see Ref.~\cite{Siemonsen:2022ivj}]. These results show that, in the most optimistic scenario, the signal would have been above the detection threshold, but does not allow us to rule out dark photon parameters. A statistically robust prediction of the expected signal from a population of Galactic BHs, to be compared with the $h_0^{95\%}$ ULs shown here, will be performed in Sec.~\ref{sec:interpretations}. The solid blue and dotted orange lines denote, respectively, the sensitivities of the narrow-band search calculated with Eq.~\eqref{eq:hsens} and the FH all-sky O3 search taken from Ref.~\cite{KAGRA:2023pio}, for reference.}
    \label{fig:search_results}
\end{figure*}

For the targets in binary systems, we discard 15 sources due to incomplete timing solutions, which do not provide all five orbital parameters necessary for the Doppler correction described in Sec.~\ref{sec:signal}~\cite{Leaci:2016oja, Leaci:2015bka}. For these cases, constructing a grid of reasonable values for the missing parameters would require ad hoc solutions, leading to either high computational costs or unreliable results~\cite{Singhal:2019dfn, Leaci:2015bka}. We did not use results from all-sky searches~\cite{LIGOScientific:2020qhb,Singh:2022hfd} as these targets have an orbital period, eccentricity or semimajor axis---when provided---outside their studied ranges. We additionally checked that, for the 3 remaining sources, the stability of the superradiance cloud would not be affected by the gravitational perturbation given by the binary companion~\cite{Berti:2019wnn, Baumann:2019ztm}. The most disruptive effect is a resonantly enhanced transition between different states of the cloud, when the binary orbital frequency matches the energy gap between two superradiance states. This is less likely to occur for a vector cloud, compared to a scalar cloud, due to additional selection rules. We checked that the orbital frequency is always smaller than the lowest possible energy splitting~\cite{Baumann:2019eav, Baumann:2019ztm}, ensuring that a disruptive resonant transition cannot have occurred yet. Therefore, we search for a CW signal from the 3 sources in a binary system with the method outlined in Sec.~\ref{sec:resamp_pipeline} and discussed in detail in Appendix~\ref{app:resamp_grid}.

Additionally, we perform a semicoherent analysis on these targets using the method from~\cite{Mirasola:2024kll}, where the coherence time is derived from the uncertainties in the binary orbital parameters, followed by a single-template search. The results, provided in Appendix~\ref{app:semi-coh_method_bin}, appear to be less sensitive than those obtained with the resampling pipeline and are therefore not included in our final results.

Lastly, 4 sources in our sample (J1748-3009, J0921-5202, J1551-0658, and J0125-2327) were previously analyzed in~\cite{LIGOScientific:2021hvc} using O2+O3 data. For these targets, we adopt the results from that study.

\subsection{Results}
\label{sec:results}
In this section, we describe the outcome of the search aimed at detecting the CW counterpart of the targets presented in Sec.~\ref{sec:targets}, applying the methods as described in Sec.~\ref{subsec:targets_and_methods}. The purpose of this analysis was to search for signals from dark photon superradiance clouds. We use LIGO data from the third observing run (O3)~\cite{KAGRA:2023pio}, specifically from the Hanford (H1) and Livingston (L1) detectors. Data from Virgo~\cite{virgo} and KAGRA~\cite{kagra} are not included, as their sensitivity curves at the relevant frequencies are worse than those of the two LIGO detectors.

No outliers were found in our analysis, allowing us to set ULs on the strain amplitude emitted by each of the targets. As detailed in Sec.~\ref{sec:methods}, the $95\%$ CL ULs were derived by injecting simulated signals into real data. Our results, compared with the largest predicted strain from a dark photon superradiance cloud~\cite{Siemonsen:2022ivj}, are shown in Fig.~\ref{fig:search_results}. Also included is a theoretical sensitivity curve for the narrow-band search, calculated as described in~\cite{AstoneColla2014,DOnofrio_2025} as
\begin{equation}
    h_{\rm sens}^{\rm NB} \sim 26 \cdot \sqrt{\frac{S_n}{T_{\rm obs}^{\rm eff}}}\,.
    \label{eq:hsens}
\end{equation}
Here, $S_n$ is the harmonic mean of the unilateral noise spectral density of the detectors, and $T_{\rm obs}^{\rm eff}$ is the observing time adjusted for the detector duty cycle, which is approximately 250 days for O3~\cite{KAGRA:2023pio}. The prefactor is computed for $10^8$ explored templates.\\

\afterpage{
\setlength{\tabcolsep}{11pt}
\renewcommand{\arraystretch}{1.22}
\LTcapwidth=\textwidth
\begin{longtable*}{cccccc}
\caption{The results of the analysis are provided for each target. We report the pulsars' name, the rotational frequencies (values are referred to the common Modified Julian Date MJD = 58700, uncertainties are omitted for simplicity), the applied methods, the equivalent coherence times, and the estimated ULs on the strain. $\Tfft^{\rm sky}$ is calculated following Appendix~D of~\cite{Mirasola:2024kll} to account for uncertainties in sky localization, as given in the ephemerides. ULs for ``all-sky'' and ``targeted'' sources are taken from~\cite{KAGRA:2023pio} and~\cite{LIGOScientific:2021hvc}, respectively. For additional parameters and uncertainties, see Tabs.\ref{tab:pos_fdot}-\ref{tab:doublets}-\ref{tab:triplets}. Targets marked with the superscript $D$ are part of doublets. Note that targets in doublets or triplets may be missing one or more companions, due to the selections described in Sec.~\ref{subsec:targets_and_methods}.}\label{tab:results}\\
\hline \hline \rule{0pt}{3ex}  
& Pulsar name (J2000) & $f_{\text{EM}}$ [Hz] & Method & $\Tfft^{\rm sky}$ [day] & $h_0^{95\%}$ \\ [1ex] 
\hline   
\endfirsthead
\multirow{16}{*}{Targets with $\dot{f}>0$} 
&J1757-2745 & 56.54 & narrow-band & 140019 & 2.6$\times 10^{-26}$ \\
&J1641+3627A & 96.36 & narrow-band & 256019 & 2.4$\times 10^{-26}$ \\
&J1748-2446C & 118.54 & narrow-band & 1056 & 3.1$\times 10^{-26}$ \\
&J1910-5959B & 119.65 & narrow-band & 21592 & 2.3$\times 10^{-26}$ \\
&J1801-0857A & 139.36 & narrow-band & 57044 & 3.4$\times 10^{-26}$ \\
&J1748-2021C & 160.59 & narrow-band & 34906 & 2.5$\times 10^{-26}$ \\
&J0024-7204C & 173.71 & narrow-band & 58172 & 2.4$\times 10^{-26}$ \\
&J1911+0101B & 185.72 & narrow-band & 7479 & 2.7$\times 10^{-26}$ \\
&J0024-7204D & 186.65 & narrow-band & 85852 & 2.6$\times 10^{-26}$ \\
&J0024-7204Z$^D$ & 219.57 & narrow-band & 2142 & 2.6$\times 10^{-26}$ \\
&J0024-7204L & 230.09 & narrow-band & 18091 & 2.7$\times 10^{-26}$ \\
&J0024-7204G$^D$ & 247.50 & narrow-band & 43902 & 2.7$\times 10^{-26}$ \\
&J1801-0857C$^D$ & 267.47 & narrow-band & 42620 & 3.8$\times 10^{-26}$ \\
&J0024-7204M$^D$ & 271.99 & narrow-band & 15324 & 2.8$\times 10^{-26}$ \\
&J1836-2354B & 309.38 & narrow-band & 453 & 4.1$\times 10^{-26}$ \\
&J0024-7204N & 327.44 & narrow-band & 14860 & 2.8$\times 10^{-26}$ \\
\hline
 \multirow{21}{*}{Targets in doublets} 
&J1748-3009 & 103.26 & targeted & ... & 7.5$\times 10^{-27}$ \\
&J0921-5202 & 103.31 & targeted & ... & 5.8$\times 10^{-27}$ \\
&J1122-3546 & 127.58 & narrow-band & ... & 2.4$\times 10^{-26}$ \\
&J1546-5925 & 128.26 & narrow-band & 32941 & 2.4$\times 10^{-26}$ \\
&J1551-0658 & 141.04 & targeted & ... & 5.7$\times 10^{-27}$ \\
&J1748-2446T & 141.15 & narrow-band & ... & 2.5$\times 10^{-26}$ \\
&J0514-4002N & 179.60 & semicoherent & 51 & 9.0$\times 10^{-26}$ \\
&J0514-4002C & 179.70 & semicoherent & 132 & 9.0$\times 10^{-26}$ \\
&J1824-2452E & 184.50 & narrow-band & ... & 2.5$\times 10^{-26}$ \\
&J1838-0022g & 196.46 & all-sky & 0.7 & 1.4$\times 10^{-25}$ \\
&J1748-2446ac & 196.58 & narrow-band & ... & 2.6$\times 10^{-26}$ \\
&J1940+26 & 207.75 & all-sky & 0.5 & 1.5$\times 10^{-25}$ \\
&J1904+0836g & 225.23 & all-sky & 0.6 & 1.5$\times 10^{-25}$ \\
&J1953+1844g & 225.23 & all-sky & 0.5 & 1.5$\times 10^{-25}$ \\
&J1326-4728E & 237.66 & narrow-band & ... & 2.9$\times 10^{-26}$ \\
&J0125-2327 & 272.08 & targeted & ... & 8.5$\times 10^{-27}$ \\
&J1844+0028g & 280.11 & all-sky & 0.4 & 1.9$\times 10^{-25}$ \\
&J1317-0157 & 343.85 & binary & ... & 2.6$\times 10^{-26}$ \\
&J0418+6635 & 343.62 & narrow-band & 1547 & 3.3$\times 10^{-26}$ \\
&J0024-7204S & 353.31 & binary & 28554 & 1.1$\times 10^{-26}$ \\
&J1308-23 & 353.40 & narrow-band & ... & 3.5$\times 10^{-26}$ \\
\hline
 \multirow{7}{*}{Targets in triplets} 
&J1342+2822F & 227.30 & narrow-band & ... & 2.8$\times 10^{-26}$ \\
&J1803-3002B & 227.43 & narrow-band & ... & 2.8$\times 10^{-26}$ \\
&J1823-3021E & 227.58 & narrow-band & ... & 2.8$\times 10^{-26}$ \\
&J0024-7204H & 311.49 & binary & 15101 & 5.2$\times 10^{-26}$ \\
&J1930+1403g & 311.53 & all-sky & 0.12 & 2.0$\times 10^{-25}$ \\
&J1624-39 & 337.84 & narrow-band & ... & 3.0$\times 10^{-26}$ \\
&J2045-68 & 337.84 & narrow-band & ... & 3.3$\times 10^{-26}$ \\
\hline\hline
\end{longtable*}
}

We want to remark that ULs on the four targeted sources from~\cite{LIGOScientific:2021hvc} shown in Fig.~\ref{fig:search_results} fall below the $h_{\rm sens}^{\rm NB}$ curve because they are not penalized by any trial factor~\cite{DOnofrio_2025,Mastrogiovanni:2017xjr}; they indeed explore a single parameter space point contrary to the methods presented in Sec.~\ref{sec:methods}.\\
For comparison, we also present the UL curve from the FH pipeline for the O3 all-sky search, as reported in~\cite{KAGRA:2022dwb}.  Additionally, all the targeted sources, along with their respective pipeline, $\Tfft^{\rm sky}$, and strain ULs, are listed in Table~\ref{tab:results}.

For all the observed targets, the reported $h_0^{95\%}$ is smaller than the maximum strain that a dark photon cloud could produce, assuming it is as young as $10^3$ yr~\cite{Siemonsen:2022ivj}, as indicated by the red stars in Fig.~\ref{fig:search_results}. This optimistic prediction generally exceeds, in most cases, also the $95\%$ CL ULs set by the all-sky search. However, it is important to note that the maximum strains shown in Fig.~\ref{fig:search_results} do not represent a statistically robust prediction of the expected signal from a population of BHs (averaging over their mass, initial spin, age, and distance distributions).

Such a detailed study was outside the scope of Ref.~\cite{Siemonsen:2022ivj}, where the search was first suggested. We attempt to fill this gap in the next section, where we provide a simple statistical framework to estimate the number of {\it observable} events--{\it i.e.} with a signal strain above $h_0^{95\%}$--from a population of Galactic BHs. An improvement in the strain sensitivity therefore directly translates into a larger number of accessible events. The formalism presented below also allows for an interpretation of the strain ULs in terms of a ``bound'' on dark photon masses and couplings, albeit valid only under the assumptions about the cloud's radio emission and the BH population properties.

\section{Interpretation of the results}
\label{sec:interpretations}

The CW search described in Sec.~\ref{sec:gw_search} resulted in ULs on the strain that can be emitted by each of the radio sources targeted here. To be compatible with observations, if these systems emit CWs at twice the frequency of the measured pulsating radio emission, the emitted strain needs to be smaller than the obtained ULs. Provided a model for the population of Galactic BHs, we can estimate how many events we would have expected with a strain above the observational threshold for a given dark photon mass and coupling. If a large number of events is expected, but none is observed, we can therefore conclude that the particular dark photon parameters are disfavored by our search.

Below, we develop an analytical method to estimate the expected number of events of CW signals at and above a given strain, given a simple model for the BH population in the Milky Way (MW). The purpose of this section is not to provide a quantitatively accurate prediction of expected events, since many of the ingredients required for the calculation are not precisely known, but to estimate whether it is reasonable to expect detectable signals in our search and future analyses. Here we use this method to interpret the results of our targeted CW search, but its applications are broader as it can also be used for all-sky and Galactic Center searches.\footnote{In this work, we focus on the signal from vector fields, but the same procedure works also for scalar fields. For the case of scalars, a similar and comprehensive study, based on Monte-Carlo simulations of galactic BHs, was done in Ref.~\cite{Zhu:2020tht}. See also~Refs.~\cite{Arvanitaki:2014wva,Baryakhtar:2017ngi, Brito:2017zvb} for analytic population estimates.}

\subsection{Galactic black hole population}
\label{sec:galactic_bh}

First, we need a model for the population of spinning BHs in the MW, since the sources targeted here---as well as the loudest signals for an all-sky search---belong to our Galaxy. The total number of BHs, as well as their spatial, age, and spin distributions are not known \textit{a priori}. Most of the existing observations rely on x-ray emission from accreting BHs in binary systems. Accurate astrometric measurements from the Gaia satellite are opening up the possibility of detecting binaries in which the BH is not interacting with its companion star, with three recent candidate events~\cite{2023AJ....166....6C, Tanikawa:2022xel, 2023MNRAS.518.1057E, 2023MNRAS.521.4323E}---including the heaviest BH of stellar origin ever observed in the MW, with a mass of $33\ M_{\odot}$~\cite{2024A&A...686L...2G}---and more expected in the upcoming data release. Even more challenging is the detection of isolated BHs, either by looking for the emission coming from accretion of the surrounding medium or by looking for photometric and astrometric gravitational microlensing events [with a recent positive detection~\cite{OGLE:2022gdj, 2022ApJ...933L..23L}]. All these measurements will help to shed light on the Galactic BH population and the mechanisms that lead to the formation of stellar-mass BHs. Lacking this information for now, we follow the existing literature and chose a model that closely follows Refs.~\cite{Zhu:2020tht,Tsuna:2018abi}. We refer the reader to those papers for a more detailed discussion on the assumptions made and simply summarize the distributions adopted below. 

The total number of BHs in the Galaxy is estimated to be of the order of $10^8$~\cite{1983bhwd.book.....S, 2017MNRAS.468.4000C}; among these, at least half of them~\cite{2013MNRAS.430.1538F} or even a dominant fraction~\cite{Wiktorowicz:2019dil, Olejak:2019pln} are expected to be isolated.
Here we assume that there are $N_{\rm BH}=10^8$ isolated BHs in the Galaxy. While this number is not known precisely, the final result is simply proportional to $N_{\rm BH}$, so it can be easily rescaled to a different normalization. In the analysis below, we exclude the four sources that are known to be in a binary system (J0024-7204H, J1748-3009, J1317-0157, and J0024-7204S).  
Accretion from the interstellar medium should not disrupt the superradiance growth or the creation of the pair plasma inside the cloud~\cite{Siemonsen:2022ivj}. Instead, this mechanism could sustain or replenish the plasma when the cloud's EM fields are too weak to initiate a cascade production of charged particles. 

For simplicity, we consider only the population of stellar mass BHs born in the Galactic disk, which we assume follows the current stellar distribution.\footnote{Signals from outside the MW, {\it i.e.}~from Andromeda, are not expected to be observable, as shown in~\cite{Zhu:2020tht} for scalar and in~\cite{Baryakhtar:2017ngi} for vector clouds.} The differential distribution of Galactic disk BHs with mass $M$, age $\tau$, spin $\chi$, at a location $\vec{r}$ can be factorized into the product of four independent distributions as
\begin{align}
    \frac{\dd n_{\rm BH}}{\dd \vec{r}\, \dd M \dd \chi \dd \tau}(\vec{r}, M, \chi, \tau) = \frac{\dd n_{\rm disk}}{\dd \vec{r}} \frac{\dd f_M}{\dd M} \frac{\dd f_\chi}{\dd \chi} \frac{\dd f_\tau}{\dd \tau}.
\end{align}
This distribution is normalized to the total number of BHs in the disk, $x_{\rm disk}N_{\rm BH}$, where the fraction of total BHs in the disk is taken to be $x_{\rm disk} = 85 \%$~\cite{Tsuna:2018abi} based on observational evidence of the distribution of the total stellar mass in the Galaxy~\cite{2015ApJ...806...96L}. 
The spatial distribution is taken to be
\begin{align}
    \label{eq:spatial_disk}
    \frac{\dd n_{\rm disk}}{\dd \vec{r}}(\rho, z) & = \frac{x_{\rm disk} \,N_{\rm BH}}{4\pi \,z_{\rm max} \,r_{d}^2}\, e^{-\rho/r_d} \,\Theta(z_{\rm max}-|z|),
\end{align}
where $\lbrace \rho, \varphi, z \rbrace$ denote the cylindrical coordinates relative to the Galactic Center. The disk's size and height are chosen to be $r_d = 2.15$ kpc and $z_{\rm max} = 75$ pc~\cite{2015ApJ...806...96L}. Strictly speaking, this is the spatial distribution of BHs at birth, while they could be moving away from their initial birthplace in case of large natal kick velocities. Here, we neglect this effect, that was included in the simulations of Refs.~\cite{Zhu:2020tht} and~\cite{Tsuna:2018abi} by tracking the BHs trajectories in the Galactic gravitational potential. Our approximation effectively introduces an error in the estimate of the BHs distances and, in turn, their observed strain amplitudes.  
For an average kick velocity of 50 km/s, most BHs tend to stay in the Galactic disk, without migrating to higher latitudes. Reference~\cite{Zhu:2020tht} showed that taking a value of 100 km/s, the expected signals were about 10\% weaker on average. This effect could be included by adopting a time-dependent spatial distribution, with older BHs allowed to extend further out in the Galaxy.
A full study of the impact of the spatial distribution on the final number of estimated events is beyond the scope of this work and is left to future studies.\footnote{Note that the simplifying assumption of the spatial distribution does not mean that the sources analyzed should follow Eq.~\eqref{eq:spatial_disk}, as the BHs can migrate to higher galactic latitudes after formation. As shown in studies of the galactic population of neutron star pulsars~\cite{Dirson:2022eyc}, the overall number of observable sources in existing catalogs can be reproduced by taking a galactic disk stellar distribution at birth similar to the one adopted here and a natal kick velocity of 75 km/s, even though the final simulated spatial distribution does not accurately reproduce the observed distribution.}

For the mass distribution, we take the Salpeter function~\cite{1955ApJ...121..161S, 2010ARA&A..48..339B} with lower and upper cutoffs $M_{\mathrm{min}}$ and $M_{\mathrm{max}}$. Additionally, we assume a flat dimensionless spin distribution between $\chi_{\rm min}$ and $\chi_{\rm max}$, and a constant BH formation rate, leading to a uniform BH age distribution between $\tau_{\rm max}$ and $\tau_{\rm min}$. Therefore we have
\begin{align} \label{eq:BH_single_dist}
    \frac{\dd f_M}{\dd M} & = \frac{1.35 }{1 - (M_{\mathrm{min}}/M_{\mathrm{max}})^{1.35}} \frac{M_{\mathrm{min}}^{1.35}}{M^{2.35}} \nonumber \\ 
    \frac{\dd f_\chi}{\dd \chi} & = \frac{1}{\chi_{\rm max}-\chi_{\rm min}}, \nonumber\\
    \frac{\dd f_\tau}{\dd \tau} & = \frac{1}{\tau_{\rm max}-\tau_{\rm min}}.
\end{align}
The maximum age of a BH is taken to be $\tau_{\rm max} = 10$ Gyr, while formation is assumed to continue until today, {\it i.e.}, $\tau_{\rm min} = 0$~\cite{2016PASA...33...24N}. The extreme values of the BHs mass and spin are varied between a few more optimistic choices and more pessimistic choices, $[M_{\rm min}, M_{\rm max}] = [5, 30], [5, 20]\ M_{\odot}$ and $[\chi_{\rm min}, \chi_{\rm max}] = [0, 1], [0, 0.5],[0, 0.3]$~\cite{Zhu:2020tht}. 

Under the above assumptions, the BH differential distribution simplifies to
\begin{equation} \label{eq:BH_dist}
      \frac{\dd n_{\rm BH}}{\dd \vec{r}\, \dd M \dd \chi \dd \tau}(\vec{r}, M) = \frac{1}{\mathcal{N}}\frac{M_{\mathrm{min}}^{1.35}}{\chi_{\rm max}\tau_{\rm max}M^{2.35}}\frac{\dd n_{\rm disk}}{\dd \vec{r}}, 
\end{equation}
where we have introduced the normalization
\begin{equation} \label{eq:BH_dist_norm}
        \mathcal{N} \equiv \frac{1}{1.35}\left[1 - \left(\frac{M_{\mathrm{min}}}{M_{\mathrm{max}}}\right)^{1.35} \right]\left( 1-\frac{\chi_{\rm min}}{\chi_{\rm max}} \right) \left(1-\frac{\tau_{\rm min}}{\tau_{\rm max}} \right)
\end{equation}
If an additional population were to be included, {\it e.g.} the BHs from the Galactic bulge, the new distribution is additive. Assuming that BHs from the Galactic bulge make up a subdominant fraction of the total BH population and that their formation stops earlier, we find that they give a negligible contribution to the total observable signals and we neglect them in our results.

\subsection{Strain distribution and expected number of events}\label{sec:CWh_dist}

In order to estimate the number of observable events with a strain above the detection threshold, we would like to obtain a density distribution for a given signal strain $h_0$, {\it i.e.} the equivalent of GW strain ``luminosity function''. For each BH with a given mass $M$, spin $\chi$, age $\tau$, and position in the Galaxy $\vec{r}$, we can compute the corresponding CW strain that an observer would see today $\bar{h}_0(M, \chi, \tau_{\rm obs}, d\ |\ \dpm)$ from Eq.~\eqref{eq:strainevo}, given a boson mass $\dpm$---which determines the GW frequency, see Eq.~\eqref{eq:obs_freq}. Here, $\tau_{\rm obs}$ denotes the time since the cloud has fully formed minus the time needed for the signal to travel from the source to the observer and is related to the BH's age as $\tau_{\rm obs} = \tau - \tau_{\rm SR}(M,\chi\ |\ \dpm) - d/c$, where $\tau_{\rm SR}$ is the superradiance growth timescale, and $d = |\vec{r}-\vec{r}_{\odot}|$ is the distance between the BH and the SSB.\footnote{For simplicity, we ignore GW and EM emission before the cloud has reached saturation, which represents a small fraction of the total lifetime of the system.} 

For any dark photon mass $\dpm$ and coupling $\varepsilon$, we can therefore obtain the density distribution of a signal of strain $h_0$ by integrating over the BHs distribution from Eq.~\eqref{eq:BH_dist} to get 
\begin{widetext}
\centering\begin{align}\label{eq:dfdlogh}
    \frac{\dd n_{h}}{\dd h_0}(h_0\ |\ \dpm, \varepsilon) & = \int \dd^3 \vec{r} \dd M \dd \chi \dd \tau \frac{\dd n_{\rm BH}}{\dd \vec{r}\, \dd M \dd \chi \dd \tau}\ \delta\left[\bar{h}_0(M, \chi, \tau_{\rm obs}, d\ |\ \dpm) - h_0\right] \Theta_{\dot{f}} \Theta_{\rm EM} \Theta_{\rm pl} \Theta_{\rm ev} \nonumber \\
    & = \frac{1}{\mathcal{N}\chi_{\rm max}\tau_{\rm max}} \int \dd M \frac{M_{\mathrm{min}}^{1.35}}{M^{2.35}} \int \dd \chi \Theta_{\rm pl} \int \dd^3 \vec{r}\, \frac{\dd n_{\rm disk}}{\dd \vec{r}}\int \dd \tau\ \delta[\bar{h}_0(M, \chi, \tau_{\rm obs}, d\ |\ \dpm) - h_0] \Theta_{\dot{f}} \Theta_{\rm EM} \Theta_{\rm ev}, 
\end{align}
\end{widetext}
where the integral over the BH's age is over the interval $\tau_{\rm SR}+d/c < \tau < \tau_{\rm max}$; {\it i.e.}, we only include BHs that have formed long enough ago to allow for the superradiance cloud to grow and the signal to travel to us. In the integral of Eq.~\eqref{eq:dfdlogh}, there are a few additional requirements that have been included in the form of 4 $\Theta$-functions, to ensure that only the signals that can actually be observed are accounted for and that the assumptions made regarding the formation and evolution of the luminous dark photon cloud hold. These conditions are as follows:
\begin{itemize}
    \item $\Theta_{\dot{f}} \equiv \Theta \left[\dot{f}_{\rm max} - \dot{f}_{\rm GW}(M, \chi, \tau_{\rm obs} |\ \dpm)\right]$ imposes an upper bound on the spin-up rate; this is required because CW searches are designed for nearly monochromatic signals which allow for small frequency drifts but lose sensitivity for larger ones. The value of $\dot{f}_{\rm max}$ is set by the specific search method and varies within the approximate range of $[2\times 10^{-15}, 2\times 10^{-11}]$ Hz/s around the measured values, for the pipelines and frequencies considered here. For the spin-up rate, we include only the intrinsic component given in Eq.~\eqref{eq:obs_fdot}\footnote{Since the intrinsic $\dot{f}$ cannot be measured, see Eq.~\eqref{eq:obs_fdot}, there could be in principle other contributions that cancel a large intrinsic spin-up rate, but we conservatively neglect such fine-tuned configurations.}. In practice, the systems with large $\dot{f}$ typically decay very quickly and do not significantly contribute to the total number of expected signals. We therefore find that the cut on $\dot{f}$ does not significantly affect the final result.
    \item $\Theta_{\rm EM} \equiv \Theta \left[F_{\rm r}(M, \chi, \tau_{\rm obs}, d\ |\ \dpm, \varepsilon) - F_{\rm th}\right] \label{eq:thetaEM}$ imposes a lower limit on the EM radio flux of the superradiance cloud, for the source to be detected by a pulsar radio survey and be a candidate for our CW search. The superradiance cloud radio flux is given by
    \begin{align} 
    \label{eq:radio_lum}
        F_{\rm r}(M, \chi, \tau, d\ |\ \dpm, \varepsilon) = f_r\frac{L(M, \chi, \tau|\ \dpm, \varepsilon)}{4\pi d^2},
    \end{align}
    where $L$ is the total cloud luminosity given in Eq.~\eqref{eq:lum} and $f_r$ is the fraction going into radio emission. As discussed in Sec.~\ref{sec:dp_superrad}, this quantity cannot be determined with current resistive force-free simulations~\cite{Siemonsen:2022ivj}. Additionally, this parameter is degenerate with the beaming fraction of the pulsating component ({\it i.e.~}the geometrical factor that takes into account that only a fraction of the sources will have a radiation beam that crosses the line-of-sight), which determines
    the fraction of sources detectable from Earth. For neutron star pulsars, the known radio beam geometry as a function of the neutron star properties, such as the rotation period, can be used to estimate the beaming fraction [see {\it e.g.} Refs.~\cite{Johnston:2020gwr, Dirson:2022eyc}]. In our case, however, the geometry of the radio emission from the cloud is not completely understood, so in principle, the beaming fraction represents another unknown free parameter. As the evidence for periodicity found in Ref.~\cite{Siemonsen:2022ivj} is mild, the beaming fraction may conservatively be zero. In this work, however, we assume a nonvanishing pulsating component exists and take $f_r = 10^{-5}$, motivated by spectra of neutron star pulsars~\cite{Kaspi:2017fwg,Manchester:2004bp}. For the detection threshold, $F_{\rm th}$, we assume the sensitivity of large-scale surveys that detected a large number of pulsars, {\it i.e.} the Parkes radio telescope in the Southern Galactic plane~\cite{2001MNRAS.328...17M} and the Arecibo telescope in the Northern plane~\cite{2006ApJ...637..446C}, both with a sensitivity of $\sim 0.15$ mJy at 1.4 GHz. For a source to be detected we require a SNR of 10, following Refs.~\cite{Johnston:2020gwr, Dirson:2022eyc}, so that $F_{\mathrm{th}} = 1.5\ {\rm mJy}$ (we ignore the gain factor coming from narrow pulse profiles that can be detected for fluxes below the nominal sensitivity). In practice, we find that most sources that have strain amplitudes above the current sensitivity, of about $10^{-26}$, are also visible with a radio survey and this condition does not significantly affect the result for $\varepsilon$ down to about $10^{-9}$ (see Sec.~\ref{sec:radio_dist} for further discussion on this point). 
    \item $\Theta_{\rm pl} \equiv \Theta\left[\tilde{\Gamma}_{e^\pm}(M, \chi\ |\dpm, \varepsilon) - 1 \right]$ requires that the electric field within the superradiance cloud is large enough to efficiently create a plasma of charged particles, which then leads to EM emission; here, $\tilde{\Gamma}_{e^\pm}$ denotes a dimensionless pair-production rate in units of the cloud's size. We refer the reader to Sec.~IV of \cite{Siemonsen:2022ivj} for a description of the processes responsible for charged particle creation within the cloud and further understanding of this condition. This requirement is relevant at small values of $\varepsilon$, and it effectively reduces to zero the number of expected events at a critical value in the range $10^{-10} - 10^{-9}$.
    \item $\Theta_{\rm ev} \equiv \Theta\left[0.1 - |M_c(\tau_{\rm obs})/M_c^{\rm true}(\tau_{\rm obs}) - 1|  \right]$ 
    imposes that the cloud decays primarily through GW emission. In practice, this requires that the cloud time evolution given by the power law from Eq.~\eqref{eq:cloudmassevo} is correct within 10\% at the time of the observation compared to the true time evolution, $M_c^{\rm true}$, that accounts for EM emission as well.\footnote{The expression for $M_c^{\rm true}$ can be obtained by solving the differential equation 
    \begin{equation}
        \frac{\dd M_c}{\dd \tau} = -\frac{M_c^2}{\tau_{\rm GW} M_s} - \frac{M_c}{\tau_{\rm EM}},
    \end{equation}
    where $\tau_{\rm GW}$ and $\tau_{\rm EM}$ are the characteristic decay timescales through GW and EM radiations, respectively, and are related to the power emitted at the cloud saturation. The solution is~\cite{Jones:2024fpg}
    \begin{equation}
        M_c^{\rm true}(\tau) = \frac{M^s_c}{e^{\tau/\tau_{\rm EM}} + \frac{\tau_{\rm EM}}{\tau_{\rm GW}}(e^{\tau/\tau_{\rm EM}}-1)},
    \end{equation}
    which reduces to Eq.~\eqref{eq:cloudmassevo} for $\tau_{\rm GW} \ll \tau_{\rm EM}$ and $\tau \ll \tau_{\rm EM}$.
    }
    This is just a self-consistency requirement to ensure the validity of the assumed time evolution. Given that when the EM emission dominates the cloud decays exponentially, rather than with a power law, it is reasonable to assume that when this occurs the contribution to the total number of signals will be negligible. However, it is possible that this assumption is overly conservative, especially at large $\varepsilon$. In any case, the decay into EM radiation, which increases for increasing $\varepsilon$, cuts off the number of expected events above a threshold in the range $10^{-7} - 10^{-6}$, in a region that is already excluded by other searches~\cite{McCarthy:2024ozh}. The effect is more pronounced for lower-frequency systems, which correspond to smaller values of $\alpha$, given the fixed range of BH masses.  
\end{itemize} 
The strain luminosity function from Eq.~\eqref{eq:dfdlogh} can be simplified using the explicit time evolution of the strain from Eq.~\eqref{eq:strainevo} and using the $\delta$-function to integrate over the BH's age (see Appendix~\ref{app:strain_dist_disk} for the full derivation). Taking the thin disk approximation, $z_{\rm max} \ll r_d$, one gets 
\begin{widetext}
        \centering
\begin{equation}\label{eq:dfdlogh_disk}
      \frac{\dd n_{h}}{\dd \log h_0} (h_0 | \dpm, \varepsilon)= \frac{x_{\rm disk }N_{\rm BH}}{2\pi \mathcal{N}} \int \dd M \frac{M_{\mathrm{min}}^{1.35}}{M^{2.35}} \int \dd \chi\ \frac{1}{\chi_{\rm max}}\frac{h_{r_d}}{h_0} \frac{\tau_{\rm GW}}{\tau_{\rm max}} \Theta_{\rm pl} \int \dd \varphi \int \dd \Tilde{\rho}\frac{\Tilde{\rho}\  e^{-\Tilde{\rho}}}{\sqrt{\Tilde{\rho}^2 + \Tilde{r}_{\rm obs}^2 - 2 \Tilde{\rho}\ \Tilde{r}_{\rm obs} \cos \varphi}}  \Theta_{\dot{f}} \Theta_{\rm EM} \Theta_{\rm ev},
\end{equation}
\end{widetext}
where we have introduced $h_{r_{d}} \equiv h^{s}_0(M, \chi, d=r_d)$, {\it i.e.} the strain emitted at cloud saturation evaluated at a distance equal to the disk's characteristic size $r_d$; additionally, we have defined $\Tilde{\rho} \equiv \rho/r_d$, $\Tilde{r}_{\rm obs} = r_{\rm obs}/r_d$, and $r_{\rm obs} \simeq 8$ kpc is the distance of the observer from the Galactic Center. The boundaries of the volume integrals have somewhat lengthy expressions and are given in Appendix~\ref{app:strain_dist_disk}.

The expected number of signals with CW strain above the observable UL $h_{0}^{95\%}$ can be computed from the cumulative distribution of the luminosity function in Eq.~\eqref{eq:dfdlogh_disk}, as
\begin{equation} \label{eq:nev_aboveUL}
    N_\mathrm{ev}(h_0>h_{0}^{95\%}\ |\ \dpm, \varepsilon) = \int_{h_{0}^{95\%}}^\infty \dd h_0\ \frac{1}{h_0}  \frac{\dd n_{h}}{\dd \log h_0}. 
\end{equation}
Note that the total number of events for an arbitrarily small observational threshold, $h_{0}^{95\%} \rightarrow 0$, does not reach the total number of BHs, $x_{\rm disk} N_{\rm BH}$. That is expected, due to the constraints given by the $\Theta$-functions in Eq.~\eqref{eq:dfdlogh_disk}. Even when these constraints are relaxed, we have $N_\mathrm{ev}(h_0>0) < x_{\rm disk} N_{\rm BH}$, since BHs that are not spinning fast enough cannot grow a superradiance cloud and will not contribute to the CW ensemble signal.
\begin{figure*}[!hpt]
    \centering 
    \includegraphics[width=0.48\textwidth]{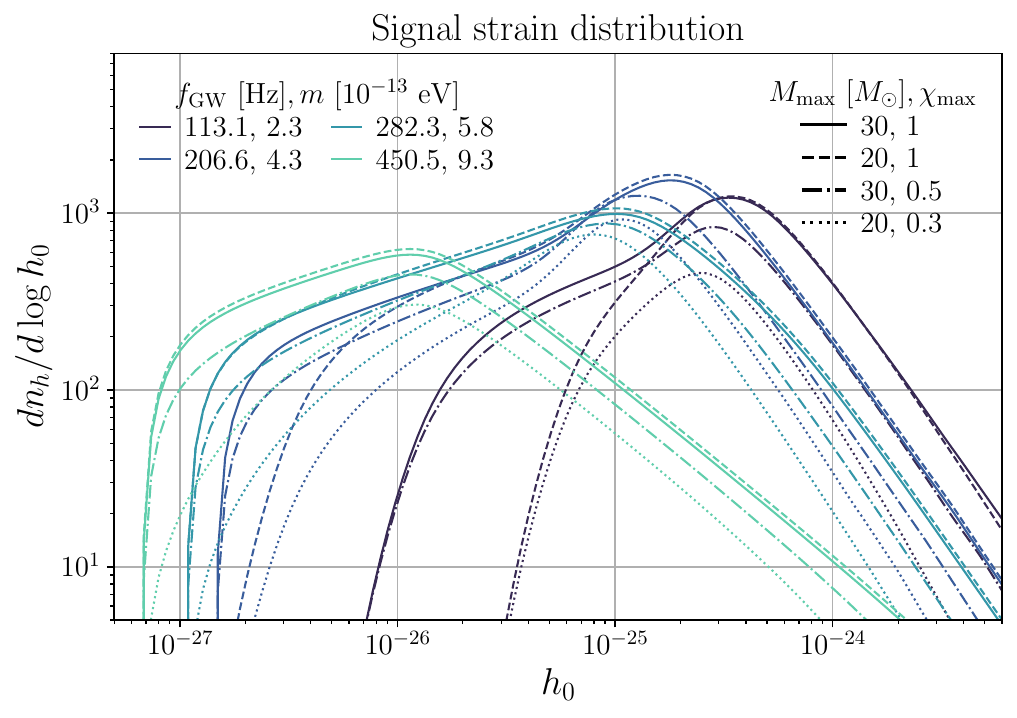}
    \includegraphics[width=0.48\textwidth]{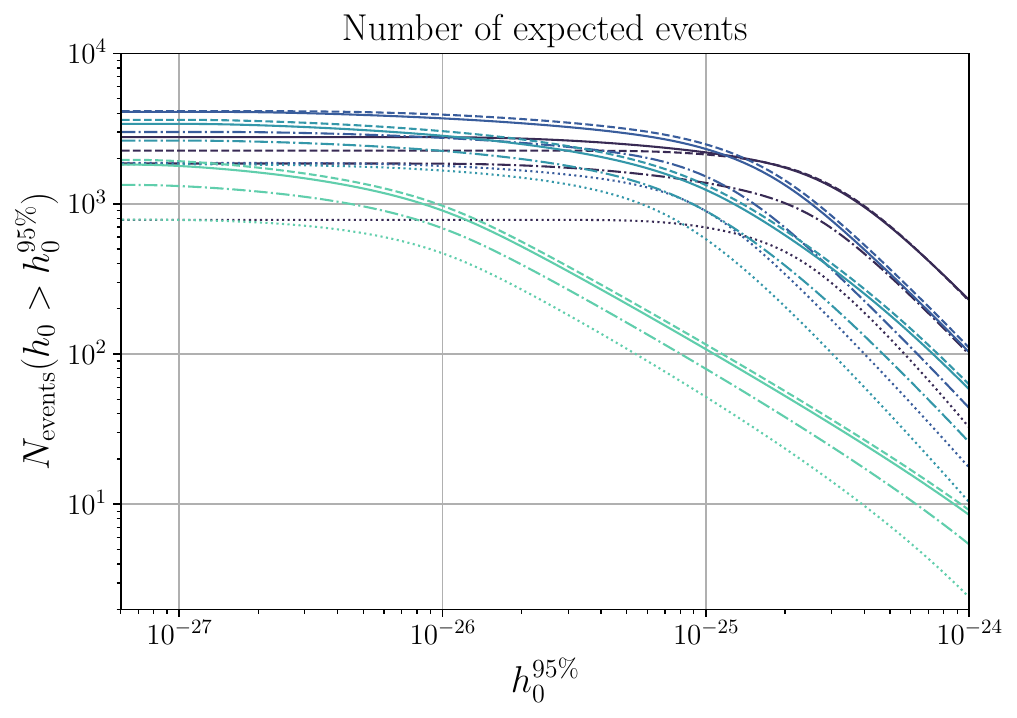}
    \caption{Signal strain density distribution of dark photon clouds around Galactic disk BHs from Eq.~\eqref{eq:dfdlogh_disk} ({\it left}) and total number of expected events with strain above observational threshold $h_{0}^{95\%}$ from Eq.~\eqref{eq:nev_aboveUL} ({\it right}). In both panels we fix the dark photon kinetic mixing coupling to $\varepsilon = 10^{-8}$ and the fraction of the cloud's radio luminosity to $f_r = 10^{-5}$ (see Fig.~\ref{fig:dndlogh_fr7} in Appendix~\ref{app:strain_dist_disk} for a comparison to the case of $f_r = 10^{-7}$). The results are shown for four representative GW frequencies corresponding to four representative pulsars from our sample of sources, with the lowest (highest) frequency shown with darker (lighter) color. The corresponding dark photon mass for each frequency is obtained as $\dpm c^2 =\pi \hbar f_{\rm GW}$ and is also indicated. The solid, dashed, dotted, and dash-dotted lines correspond to different assumptions about the Galactic BH population parameters, varying the maximum possible BH mass and spin, as labeled (see Sec.~\ref{sec:galactic_bh}).}\label{fig:dndlogh}
\end{figure*}
Here we use the expressions above to look at the strain luminosity function and expected number of events at the GW frequencies for the sources targeted in the analysis of Sec.~\ref{sec:gw_search}. To this end, there is one more caveat that needs to be addressed: the estimates above give the events distribution for a given {\it boson mass}, while our search targets specific {\it GW frequencies}. As discussed in Sec.~\ref{sec:targets}---see Eq.~\eqref{eq:obs_freq}---the observed GW frequency is fixed by the dark photon mass, up to a Doppler correction that varies for each system and that leads to a spread of $v_{\rm rad}/c \sim \mathcal{O}(10^{-3})$~\cite{Zhu:2020tht}. This means that the expected events will not all be clustered at exactly the same frequency $\left. f_{\rm GW}\right|_{\rm obs} = \dpm c^2/(\pi \hbar)$. However, this is irrelevant for both the targeted and the blind searches for the following reason: when using a radio target, the {\it observed} frequency (including the Doppler shift) is known for each source, while for a blind search a wide range of frequencies is scanned over. In both cases, then, the signal cannot be missed.\footnote{CW searches have been shown to be robust even in presence of moderate-to-large signal clustering due to many and strong CW sources~\cite{Pierini:2022wgc}.} Since the strain sensitivity varies mildly between nearby frequency bins, the estimates of the expected events made here should still hold. In the results below, we therefore fix $\dpm c^2 = 2\pi \hbar \left. f_{\rm EM}\right|_{\rm obs}=\pi \hbar \left. f_{\rm GW}\right|_{\rm obs}$ to infer the values of dark photon masses to which our search is sensitive. However, it must be noted that the exact value of the mass is unknown within a factor of about $v_{\rm rad}/c \sim \mathcal{O}(10^{-3})$.

We solve the integrals numerically and use the \texttt{SuperRad} code developed in Ref.~\cite{Siemonsen:2022yyf} to compute the growth and decay time scales, the spin-up rate, the cloud's mass, and the characteristic strain. In the small $\alpha$ limit, these quantities reduce to the analytical expressions given in Sec.~\ref{sec:dp_superrad}.\footnote{In the case of vector field superradiance the systems with large gravitational coupling $\alpha$ are very short-lived and do not contribute significantly to the total observable signals; therefore we can safely work in the small-$\alpha$ approximation.} The strain distribution for the disk component of the Galactic BHs population from Eq.~\eqref{eq:dfdlogh_disk} is shown in the left panel of Fig.~\ref{fig:dndlogh} for $\varepsilon = 10^{-8}$ and four different representative CW frequencies from the sample of sources analyzed here, corresponding to four different values of the boson mass. We derive the distribution for four sets of assumptions about the BH population parameters, by varying the maximum mass and initial spin in Eqs.~\eqref{eq:BH_single_dist}. Note that the result has a mild dependence on the maximum allowed BH spin. Taking $\chi_{\rm max} = 0.5$ reduces the number of events only by a few tens of percent compared to $\chi_{\rm max} = 1$, instead of an order of magnitude as in the case of scalar superradiance clouds~\cite{Zhu:2020tht}. This is expected since the GW emission from vector clouds is stronger than for scalar ones, and this allows for observable signals even from slowly spinning BHs. In Appendix~\ref{app:strain_dist_disk} we show the strain distribution and the number of observable events for a fraction of radio emission of the superradiance cloud of $f_r = 10^{-7}$ (see Fig.~\ref{fig:dndlogh_fr7}); compared to $f_r = 10^{-5}$ considered here, the lower value only moderately reduces the total number of events for strain amplitudes below $\sim 10^{-26}$. We checked that choosing a higher value of $f_r = 10^{-3}$ gives the same results shown in Fig.~\ref{fig:dndlogh}.

The shape of the distribution in Fig.~\ref{fig:dndlogh} is strongly affected by the $\varepsilon$-dependent conditions $\Theta_{\rm pl}$ and $\Theta_{\rm ev}$. In particular the position of the peak at a critical strain amplitude in the distribution is dictated by $\Theta_{\rm ev}$, which shuts off the contributions from systems with a small GW signal (small $\alpha$ coupling). 

The expected number of observable events from Eq.~\eqref{eq:nev_aboveUL} is shown as a function of an arbitrary minimum strain sensitivity $h_{0}^{95\%}$ in the right panel of Fig.~\ref{fig:dndlogh}. Since the luminosity function peaks at a given characteristic strain, which depends on the source frequency (boson mass), the number of events plateaus below this characteristic strain. For the value of $\varepsilon = 10^{-8}$ shown in Fig.~\ref{fig:dndlogh}, the lowest frequency systems peak at values of $h$ between $10^{-26}$ and $10^{-25}$, {\it i.e.} strain amplitudes above the current sensitivity of a fully coherent search, but below the sensitivity of an all-sky search. In this case, the search performed here gives a significantly higher chance of detection compared to the blind searches done before. For the highest frequency systems, on the other hand, the peak is still below the current most sensitive targeted searches. Notice that the position of the peak moves to higher (lower) strains for higher (lower) values of $\varepsilon$, for all systems. Therefore future searches will improve in particular in the low coupling, large boson mass regime. 

\begin{figure*}[!hpt]
    \centering 
    \includegraphics[width=0.48\textwidth]{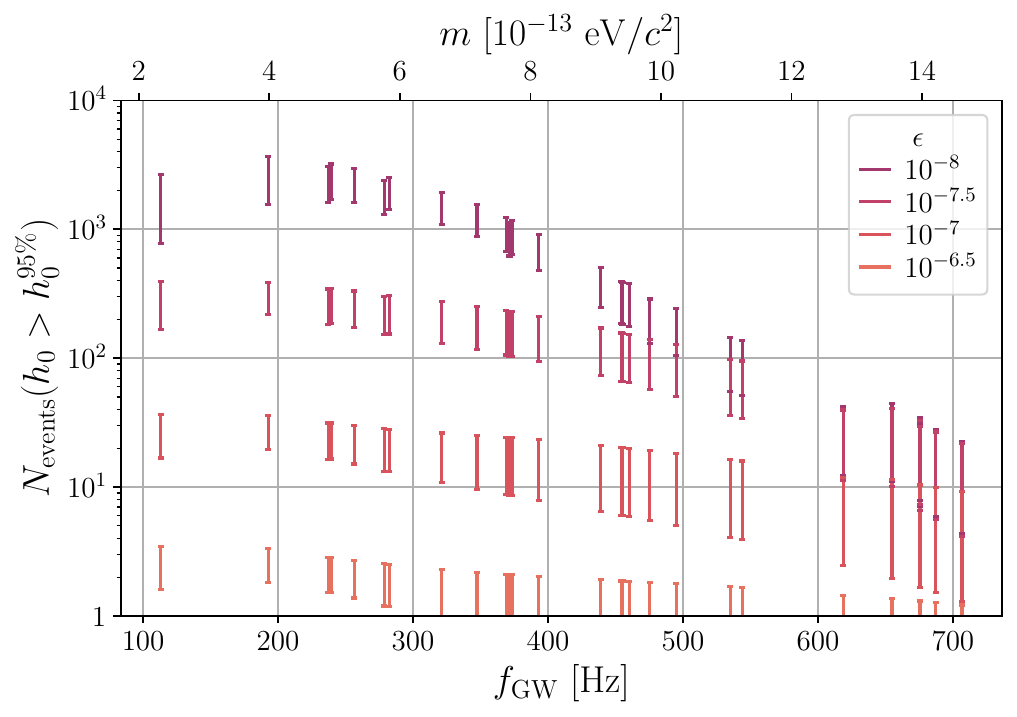}
    \includegraphics[width=0.48\textwidth]{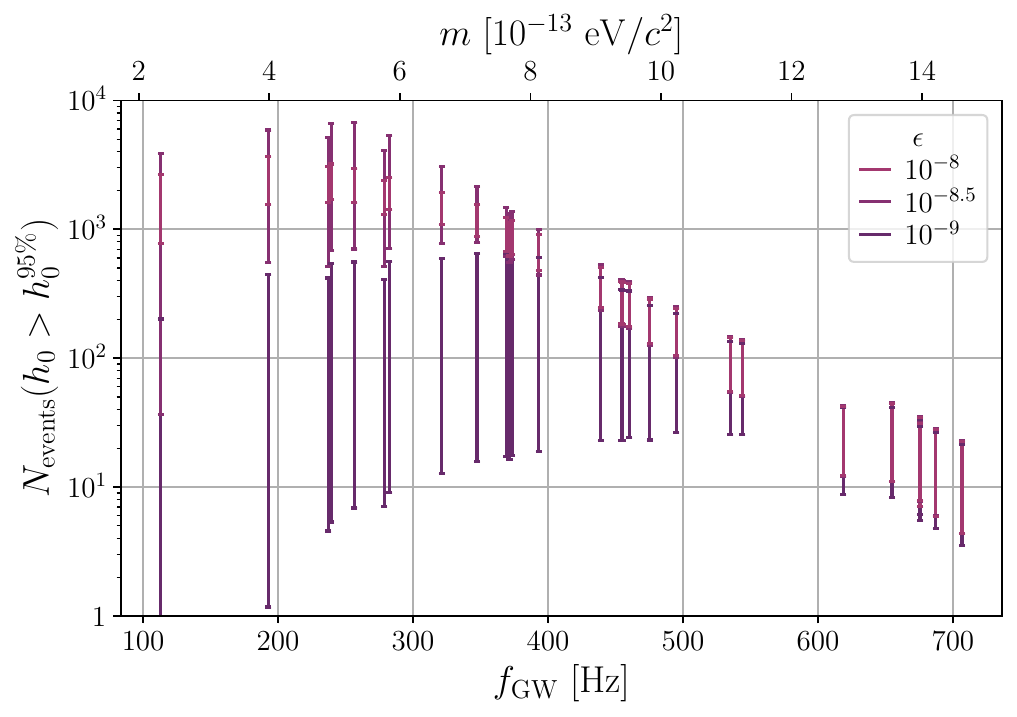}
    \caption{Number of expected events above the strain ULs obtained in this work and reported in Table~\ref{tab:results}, as a function of the source CW frequency (bottom x-axis) and corresponding dark photon mass ($\dpm c^2 =\pi \hbar f_{\rm GW}$, top x-axis). Only the sources that were analyzed with the narrow-band pipeline are included, for clarity; the result for the remaining sources is shown in Fig.~\ref{fig:Nev_targets_all} in Appendix~\ref{app:Nevents_all}. The error bar denotes the uncertainty coming from the unknown properties of the BH disk population and is computed by taking the highest and lowest number of events obtained when varying the maximum mass and spin of the BH distribution as explained in Sec.~\ref{sec:galactic_bh} and shown in Fig.~\ref{fig:dndlogh}. Lighter (darker) colors correspond to larger (smaller) values of the dark photon kinetic mixing parameter $\varepsilon$. Notice that the number of events peaks for $\varepsilon \sim 10^{-8}$ and decreases at both larger (left panel) and smaller (right panel) couplings--see the text for more details. The results are obtained by fixing the fraction of the cloud's radio luminosity to $f_r = 10^{-5}$ (see the text and Appendix~\ref{app:strain_dist_disk} for a discussion on the dependence on this parameter).
    }\label{fig:Nev_targets}
\end{figure*}

In Fig.~\ref{fig:Nev_targets}, we show the expected number of events for the sources targeted by the ``narrow-band'' method in this work (see Sec.~\ref{subsec:narrowband}), using the $h_{0}^{95\%}$ derived in the analysis presented in Sec.~\ref{sec:gw_search} and reported in Table~\ref{tab:results} (results for the sources analyses with the other pipelines are shown in Fig.~\ref{fig:Nev_targets_all} in Appendix~\ref{app:Nevents_all}\footnote{We choose to present only the sources targeted by the ``narrow-band'' method purely to allow for a clearer visualization of the results. Fluctuations in the number of events from one source to the other due to fluctuations in $h_{0}^{95\%}$ coming from different sensitivities from different analysis pipelines would make the already busy figure hard to read.}). The number of events is shown for a few different choices of the kinetic mixing coupling $\varepsilon$ and including the uncertainties from varying the BH population parameters. The results show that for most of the sources, at least $10-100$ events should have been above the threshold for a wide range of couplings, between about $10^{-9} \lesssim \varepsilon \lesssim 10^{-7}$. For smaller couplings, most superradiance clouds never produce a plasma of charged particles to emit EM radiation. For large couplings, on the other hand, most systems would start decaying through EM emission at some point during their evolution and decay exponentially fast; we stress that the latter behavior is not captured by our treatment and it is therefore possible that our estimates are overly pessimistic at larger values of $\varepsilon$.

\begin{figure*}[t]
    \centering 
    \includegraphics[width=0.48\textwidth]{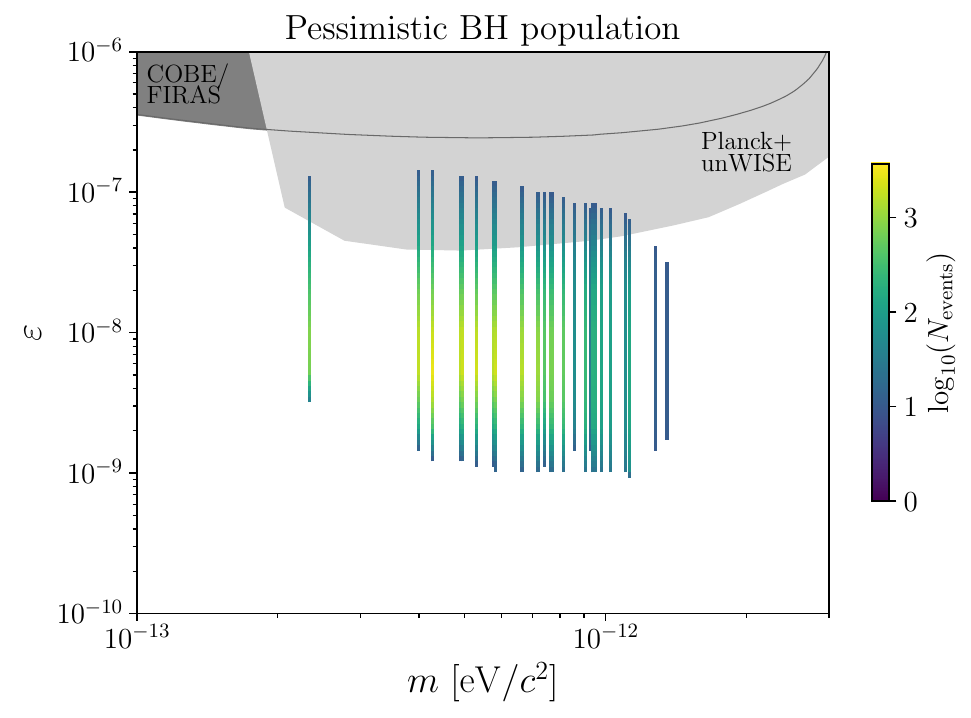}
    \includegraphics[width=0.48\textwidth]{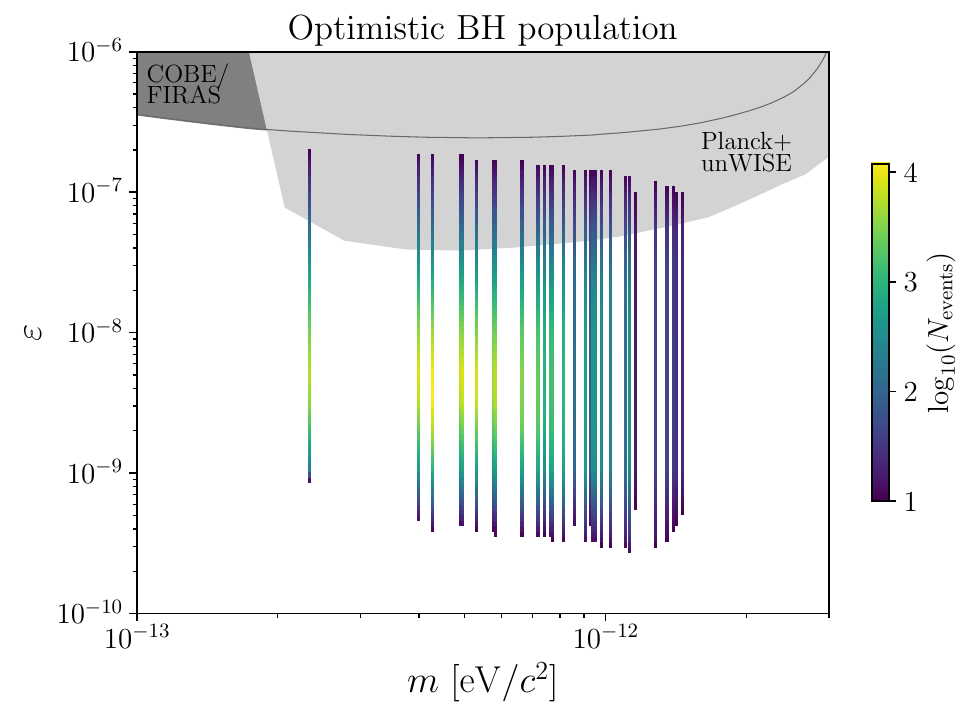}
    \caption{Regions of the dark photon parameter space where we would have expected at least 10 events above the $h_{0}^{95\%}$ strain ULs derived in this analysis (see Sec.~\ref{sec:gw_search}, Table~\ref{tab:results}). Each vertical line corresponds to one pulsating radio source, whose frequency determines the accessible dark photon mass, $\dpm c^2 =\pi \hbar f_{\rm GW}$, up to an unknown Doppler correction of $v_{\rm rad}/c \sim \mathcal{O}(10^{-3})$--see text for further discussion (the width of the lines is chosen for visualization purposes only and does not correspond to an uncertainty on the value of the mass). The color in each line is set by the number of expected observable events. We conservatively set a threshold of $N_{\rm events} = 10$ to {\it disfavor} the corresponding dark photon parameters. The left (right) panel corresponds to a choice of pessimistic (optimistic) parameters for the Galactic BH population, {\it i.e.},~taking the lowest (highest) number of events obtained when varying the maximum mass and spin of the BH distribution as explained in Sec.~\ref{sec:galactic_bh} (spread shown by the bars in Fig.~\ref{fig:Nev_targets}). Notice that the color bar range is different in the left and right panels. The results are obtained by fixing the fraction of the cloud's radio luminosity to $f_r = 10^{-5}$ (see text and Appendix~\ref{app:strain_dist_disk} for a discussion on the dependence on this parameter). Existing bounds are reproduced from Ref.~\cite{AxionLimits} and come for cosmic microwave background (CMB) photon conversion into dark photons, leading to spectral distortions in the CMB black body spectrum measured by COBE/FIRAS~\cite{Mirizzi:2009iz, Caputo:2020bdy} (dark gray) and cross-correlation between the CMB temperature anisotropies measured by Planck and the unWISE galaxies~\cite{McCarthy:2024ozh} (light gray).}
    \label{fig:dark_photon}
\end{figure*}

Finally, the number of observable events as a function of dark photon parameters is shown in Fig.~\ref{fig:dark_photon}. We define a point in the parameter space to be ``observable'' if at least 10 events are expected for that values of mass and coupling, setting a conservative threshold given the uncertainties related to the unknown BH populations. The results show how there is a wide range of unexplored parameters where our search could have produced an observable signal. We stress, however, that the color bands in the figure should not be interpreted as a robust exclusion bound of dark photon masses and couplings, due to the assumptions made on the properties of the EM emission of the superradiance cloud, which cannot be definitively inferred from existing simulations~\cite{Siemonsen:2022ivj}. At most frequencies, the current LVK sensitivity already has access to at least tens of events, even assuming moderately spinning and light BHs. A future, more sensitive, search would allow to improve in particular at high frequencies.

The expressions derived in this section to compute the number of events, Eqs.~\eqref{eq:dfdlogh_disk} and \eqref{eq:nev_aboveUL}, and the results presented above depend on two sets of assumptions: the properties of the BH population and the properties of the cloud's EM emission. A similar analysis could be repeated for blind all-sky CW searches; this case would only depend on the BH population assumptions and would be independent of the dark photon EM coupling (assuming  $\varepsilon$ small enough for GW emission to dominate the cloud evolution). One could therefore derive a ``bound'' that applies in the whole range of frequencies (dark photon masses) covered by CW searches and roughly for all values of $\varepsilon$ below $\sim 10^{-7}$; we leave a detailed study to future work.

Since the dark photon superradiance cloud emits EM radiation, one could ask the question of whether these objects would not first be seen as an anomaly in radio surveys, instead of a CW search. In the next section, we estimate the expected number of superradiance clouds with a radio flux above the sensitivity of a typical radio telescope, following a similar procedure as the one outlined above.

\subsection{Radio flux distribution and expected number of dark photon pulsars} \label{sec:radio_dist}

Using the approach outlined in the previous section, we can derive a probability distribution for the radio flux density $F_r$ seen by a radio survey for the Galactic BH population due to the dark photon superradiance EM emission (see Sec.~\ref{sec:dp_superrad}). Considering only the dominant disk component of the BH population and taking the thin disk approximation $z_{\rm max} \ll r_d$ and $\tau_{\rm min} = 0$, we get (see Appendix~\ref{app:flux_dist_disk} for the full derivation)
\begin{widetext}
        \centering
\begin{equation}\label{eq:dfdlogF_disk}
     \frac{\dd n_{F_r}}{\dd \log F_r}(F_r | \dpm, \varepsilon) = \frac{x_{\rm disk }N_{\rm BH}}{2\pi \mathcal{N}} \int \dd M \frac{M_{\mathrm{min}}^{1.35}}{M^{2.35}} \int \dd \chi\ \frac{1}{\chi_{\rm max}}\frac{f_r \varepsilon^2 L^s}{4\pi r_d^2 F_r} \frac{\tau_{\rm GW}}{\tau_{\rm max}} \Theta_{\rm pl} \int \dd \varphi \int \dd \Tilde{\rho}\ \frac{\Tilde{\rho}  e^{-\Tilde{\rho}}}{\Tilde{\rho}^2 + \Tilde{r}_{\rm obs}^2 - 2 \Tilde{\rho}\ \Tilde{r}_{\rm obs} \cos \varphi}  \Theta_{\rm ev}, 
\end{equation}
\end{widetext}
where $L^s$ is the EM luminosity emitted at cloud saturation given by taking $M_c = M_c^s$ in Eq.~\eqref{eq:lum}. The cloud radio flux in terms of the EM luminosity is given by Eq.~\eqref{eq:radio_lum} and all the other quantities were defined in Sec.~\ref{sec:CWh_dist}. The boundaries of the volume integral are given in Appendix~\ref{app:flux_dist_disk}.

The expected number of visible sources with a flux density above an observable threshold $F_{r, 0}$ can be computed from the cumulative distribution of the luminosity function from Eq.~\eqref{eq:dfdlogF_disk} as
\begin{equation} \label{eq:nsources_aboveUL}
    N_\mathrm{sources}(F>F_{r, 0}\ |\ \dpm, \varepsilon) = \int_{F_{r, 0}}^\infty \dd F_r\ \frac{1}{F_r} \frac{\dd n_{F_r}}{\dd \log F_r}. 
\end{equation}
For large-scale radio surveys, like the Parkes radio telescope in the Southern Galactic plane~\cite{2001MNRAS.328...17M} and the Arecibo telescope in the Northern plane~\cite{2006ApJ...637..446C}, the detection threshold is of the order of $1.5\ {\rm mJy}$ at 1.4 GHz (requiring a SNR of 10; this is the threshold that we assumed in Sec.~\ref{sec:CWh_dist} for the condition $\Theta_{\rm EM}$). 

\begin{figure*}[!ht]
    \centering 
    \includegraphics[width=0.48\textwidth]{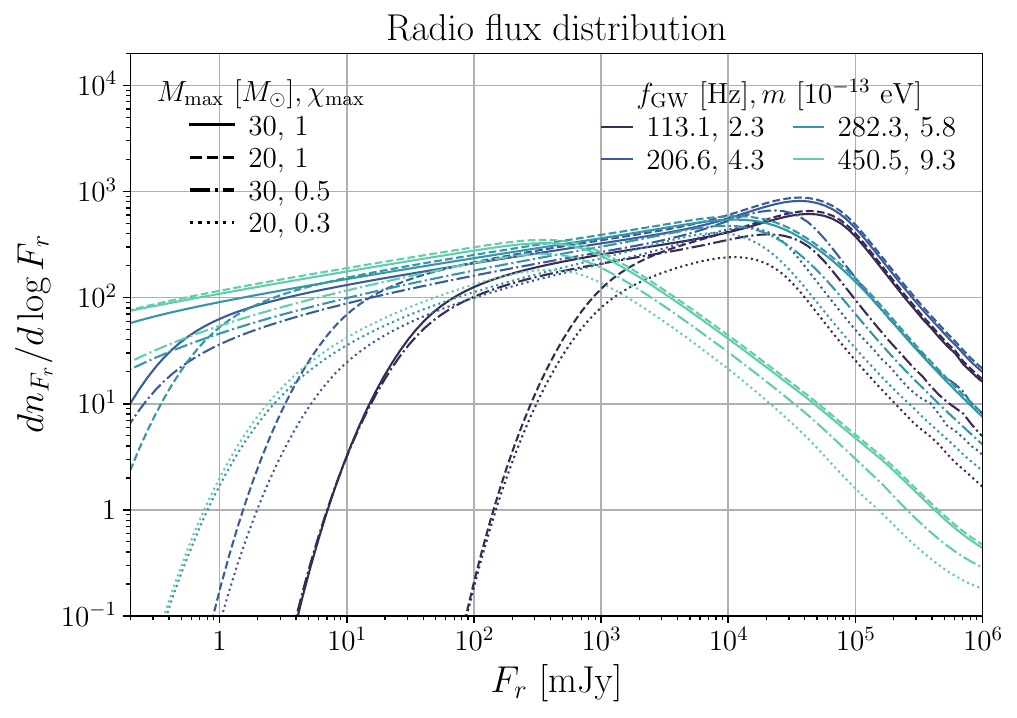}
    \includegraphics[width=0.48\textwidth]{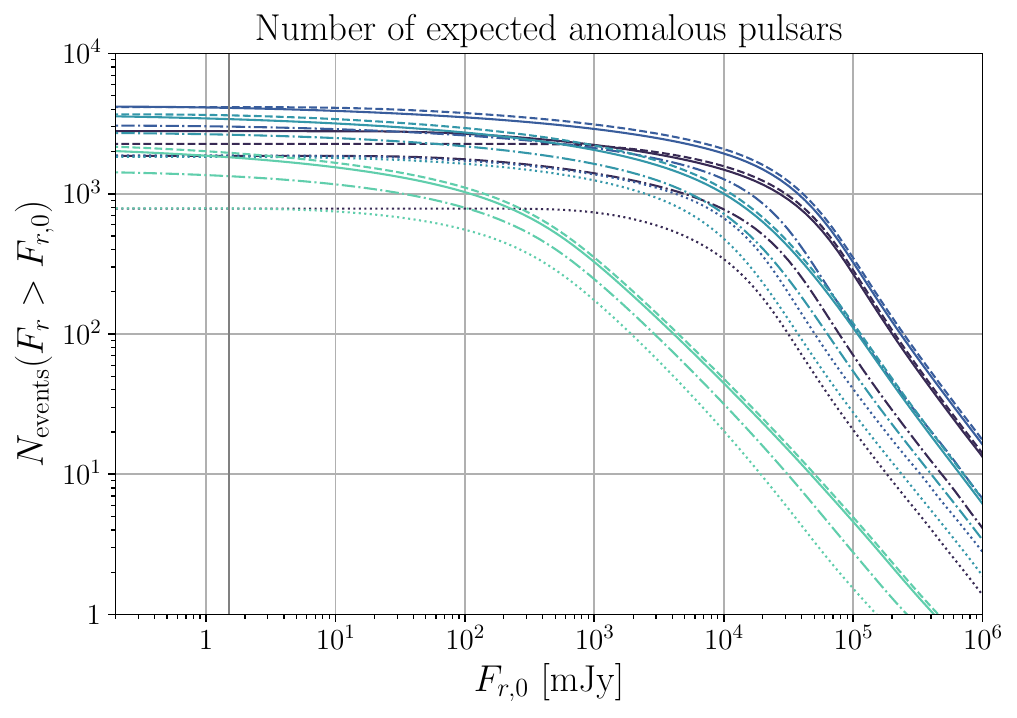}
    \caption{Radio flux density distribution of dark photon clouds around disk Galactic BHs from Eq.~\eqref{eq:dfdlogF_disk} ({\it left}) and total number of observable anomalous pulsating sources above observational threshold $F_{\rm th}$, from Eq.~\eqref{eq:nsources_aboveUL} ({\it right}). In both panels, we fix the dark photon kinetic mixing coupling to $\varepsilon = 10^{-8}$ and the fraction of the cloud's radio luminosity to $f_r = 10^{-5}$ (see the text for an understanding of the dependence on this parameter). The results are shown for four representative GW frequencies corresponding to four representative pulsars from our sample of sources, with the lowest (highest) frequency shown with darker (lighter) color. The corresponding dark photon mass for each frequency is obtained as $mc^2 = \pi \hbar f_{\rm GW}$ and is also indicated. The solid, dashed, dotted, and dash-dotted lines correspond to different assumptions about the Galactic BH population parameters, varying the maximum possible BH mass and spin, as labeled (see Sec.~\ref{sec:galactic_bh}). The vertical gray line at $F_{r, 0} = 1.5$ mJy represents the threshold sensitivity of large-scale radio pulsar surveys that we assume in this work.
    }\label{fig:dfdlogF}
\end{figure*}

The radio flux distribution for the BHs in the Galactic disk from Eq.~\eqref{eq:dfdlogF_disk} and the number of expected events from Eq.~\eqref{eq:nsources_aboveUL} are shown in the left panel of Fig.~\ref{fig:dfdlogF}, for $\varepsilon = 10^{-8}$ and $f_r = 10^{-5}$. We show the result for four different representative CW frequencies from the sample of sources analyzed in this work and four sets of assumptions about the BH population parameters, by varying the maximum mass and initial spin in Eqs.~\eqref{eq:BH_single_dist}. The result only depends on the ratio $f_r/F_r$; therefore, a rescaling to higher (lower) values of $f_r$ would correspond to a shift of the distribution towards higher (lower) flux densities $F_r$ ({\it i.e.},~a rescaling of the x-axis in Fig.~\ref{fig:dfdlogF}). Notice that the curves on the right panel of Fig.~\ref{fig:dfdlogF} asymptote to values of $N_{\rm events}$ for $F_{r, 0} \rightarrow 0$ which is equal to the asymptotic value of $N_{\rm events}$ for $h_{0} \rightarrow 0$ from Fig.~\ref{fig:dndlogh}. This is expected since, for an arbitrarily low detection threshold, we would be able to see both CW and radio signals from all the Galactic BHs with a luminous superradiance cloud.

From the right panel of Fig.~\ref{fig:dfdlogF} we can see that given our radio flux and BH population assumptions, $\sim 10^3$ anomalous pulsars with nearly coincident frequencies are within the observational reach of radio surveys for $\varepsilon = 10^{-8}$ and for rotational frequencies (and vector boson masses) in the $\sim 0.1$~kHz range. In fact, for the aforementioned parameters, current radio surveys already have the potential to observe \textit{all} the luminous superradiance clouds within our Galaxy.

In the catalog investigated in \cite{Siemonsen:2022ivj} and in this paper, only pairs or triplets of same-frequency anomalous radio sources have been found, which for $\varepsilon = 10^{-8}$, disfavors radio flux fractions of the order of $f_r \sim 10^{-5}$, and instead suggests that the cloud's pulsating radio emission efficiency could be very small, of the order of $f_r \sim 10^{-9}$. If this is correct, one would indeed expect to see only a few frequency doubles or triplets in current pulsar catalogs, and as the sensitivity and coverage of the radio surveys increase, additional anomalous sources in the same radio frequency bins should accumulate. This also highlights the importance of performing follow-up searches of GW emission from such anomalous sources, as these can provide a confirmation of the superradiant hypothesis even when the number of observed radio sources is small.

\section{Conclusions}\label{sec:conclusion}

We have performed the first search for CW signals from BH superradiance clouds of dark photons that are coupled through a kinetic mixing to the Standard Model electromagnetic sector. In the presence of this coupling, superradiance clouds around spinning BHs in the Galaxy could manifest as ``anomalous'' pulsating radio sources, which have almost coincident rotational periods---set by the boson's mass---and are spinning-up over time. Currently, we do not see an excess of ``anomalous'' pulsars in existing pulsar catalogs, but a few sources with coincident frequency and observed spin-up rates are found. A detection of continuous, nearly monochromatic GWs signals from these systems would constitute a smoking-gun signature of superradiance. Here we perform the first search for these signals for 34 new sources.

Each target is analyzed with a tailored pipeline that accounts for the uncertainties on the measured ephemerides. We adopt two fully coherent methods (narrow-band and resampling) and one semicoherent method to study both isolated sources and sources that are in a binary system.
We do not find any statistically significant outlier and obtain ULs on the strain amplitude of each target through the injection of fake signals in real data. Our results, at the 95\% CL, range from $1.05\times 10^{-26}$ for J0024-7204S to $9\times 10^{-26}$ for J0514-4002N.

We use these results, together with existing upper limits from previous analyses of 10 additional sources, to estimate the region of dark photon parameter space probed by the corresponding CW searches. Given a population of $8.5 \times 10^{7}$ isolated BHs in Galactic disk, we estimate that as many as about 3600 signals (or 11800 for optimistic BHs population parameters) should have been above the detection threshold, for dark photon masses around $5\times 10^{-13}$ and $\varepsilon \sim 5 \times 10^{-9}$. These numbers reduce to about 10 signals at the extreme values of the dark photon parameter space (see Fig.~\ref{fig:dark_photon}), covering a discrete set of masses between about $2\times 10^{-13}$ $\rm{eV}/c^2$ and $1.4\times 10^{-12}$ $\rm{eV}/c^2$ and down to kinetic mixing couplings of $10^{-9}$, nearly two orders of magnitude below the strongest existing CMB bounds~\cite{McCarthy:2024ozh}. Future, more sensitive CW searches can improve the reach, especially at high masses (high GW frequencies) and low couplings where most of the signals are still below the current detection threshold. 
In obtaining these results, we make multiple assumptions that require further confirmation both from observations and theoretical or numerical studies. 

The signal features depend on properties of the superradiating BHs, as is the case for other superradiance observational signatures. In our case, poor understanding of the overall number, spatial, age, mass, and spin distributions of Galactic BHs limits our ability to predict the superradiance ensemble signal. To better quantify these uncertainties, one could perform a more comprehensive study testing simulation-based distributions of Galactic BHs, using results from recent Galactic simulations such as ~\cite{Olejak:2019pln, Sweeney:2022fxx}. Ultimately, simulation results will have to be validated against population studies inferred from LVK merger events and future precise astrometric measurements. Our search is additionally based on the assumption that superradiance clouds emit pulsating radio signals, at a frequency given by the rotational frequency of the cloud and half the CW frequency. This assumption is motivated by the analogy between the plasma-filled dark photon cloud and a neutron star magnetosphere and is supported by the mild evidence for periodicity found in Ref.~\cite{Siemonsen:2022ivj}. We find that the prediction should not change significantly if the pulsating radio emission is at least  a fraction of $f_r \simeq 10^{-7}$ of the total cloud's EM emission (see Secs.~\ref{sec:CWh_dist},~\ref{sec:radio_dist}, and Appendix~\ref{app:strain_dist_disk} for further discussion on the dependence on this parameter).

While in this work we mostly focus on performing GW follow-up searches of pulsating EM sources, in Sec.~\ref{sec:interpretations} we also find that EM searches alone could be used to probe dark photon superradiance, since a large number of sources is expected to be above the detection threshold of existing radio surveys. It could be inferred that the whole range of dark photon masses and a similar range of $\varepsilon$ as shown in Fig.~\ref{fig:dark_photon} is disfavored by the currently observed radio sources. We stress, however, that the potential of excluding dark photon parameter space for searches relying on the radio signal alone is limited by the strong assumption on the pulsating component of the cloud emission, that cannot be robustly predicted at present. The goal of this work was the search of a CW counterpart from promising candidate sources. In order to interpret the result of the search, we also develop a method to potentially exclude dark photons. Due to the associated uncertainties discussed in Sec.~\ref{sec:interpretations}, however, the resulting constraints are not fully robust. Since dark photon clouds are expected to have a high-energy counterpart, the same sources could be seen by large-scale x-ray surveys like Swift-XRT \cite{Burrows:2005gfa} and eROSITA \cite{Predehl:2017}. A combination of low- and high-energy observations could be used to rule out the existence of these objects or identify potential dark photon cloud candidates. Ultimately, however, a detection of a CW counterpart would be required to confirm the superradiance origin.

This work paves the way for a new class of dark photon superradiance searches, that combines GW and EM observations. This strategy complements other probes of BH superradiance that are based on GW observations and spin population measurements and target both scalar and vector bosons. Planned upgrades of existing GW detectors \cite{KAGRA:2013rdx}, as well as future instruments [ET \cite{Punturo:2010zz,Hild:2010id}, CE \cite{Reitze:2019iox,Evans:2021gyd}, LISA \cite{amaroseoane2017laserinterferometerspaceantenna}, DECIGO \cite{Kawamura:2011zz}, LGWA \cite{Ajith:2024mie}], will offer improved sensitivity over complementary frequency intervals, ranging from $\sim 10^{-4}$ to $\sim 10^3$ Hz. These future observatories will be able to cover a wide boson mass range, from $\sim 10^{-18}$ to $10^{-11}$ $\rm{eV}/c^2$, providing an exciting opportunity to discover a signal from new ultralight boson clouds.

\begin{acknowledgments}
This research has made use of data or software obtained from the Gravitational Wave Open Science Center~\cite{gwosc}, a service of the LIGO Scientific Collaboration, the Virgo Collaboration, and KAGRA. This material is based upon work supported by NSF's LIGO Laboratory which is a major facility fully funded by the National Science Foundation, as well as the Science and Technology Facilities Council (STFC) of the United Kingdom, the Max-Planck-Society (MPS), and the State of Niedersachsen/Germany for support of the construction of Advanced LIGO and construction and operation of the GEO600 detector. Additional support for Advanced LIGO was provided by the Australian Research Council. Virgo is funded, through the European Gravitational Observatory (EGO), by the French Centre National de Recherche Scientifique (CNRS), the Italian Istituto Nazionale di Fisica Nucleare (INFN) and the Dutch Nikhef, with contributions by institutions from Belgium, Germany, Greece, Hungary, Ireland, Japan, Monaco, Poland, Portugal, Spain. KAGRA is supported by Ministry of Education, Culture, Sports, Science and Technology (MEXT), Japan Society for the Promotion of Science (JSPS) in Japan; National Research Foundation (NRF) and Ministry of Science and ICT (MSIT) in Korea; Academia Sinica (AS) and National Science and Technology Council (NSTC) in Taiwan.

We would like to acknowledge CNAF for providing the computational resources.

We would like to also acknowledge the ATNF Pulsar Catalog, Manchester \textit{et al.}~\cite{Manchester:2004bp}, see Ref.~\cite{pulsarlist} for updated versions.

We thank the Institute for Nuclear Theory at the University of Washington for its kind hospitality and stimulating research environment. This research was supported in part by the INT's U.S. Department of Energy Grant No. DE-FG02-00ER41132.

M.B. and C.M. would like to thank Sylvia Zhu and Maria Alessandra Papa for helpful conversations. M.B. is supported by the U.S. Department of Energy Office of Science under Award No. DE-SC0024375 and the Department of Physics and the College of Arts and Sciences at the University of Washington. M. B. is also grateful for the hospitality of Perimeter Institute, where part of this work was carried out. This work was also supported by a grant from the Simons Foundation (No. 1034867, Dittrich).

Research at Perimeter Institute is supported in part by the Government of Canada through the Department of Innovation, Science and Economic Development Canada and by the Province of Ontario through the Ministry of Colleges and Universities.
W.E. acknowledges support from a Natural Sciences and Engineering Research
Council of Canada Discovery Grant and an Ontario Ministry of Colleges and
Universities Early Researcher Award. 
\end{acknowledgments}

\section{Data availability}
The data that support the findings of this article are openly available~\cite{KAGRA:2023pio}.

\appendix

\setcounter{tocdepth}{1}

\section{Binary targets \label{app:resamp_grid}}

We applied the method described in Sec.~\ref{sec:resamp_pipeline} to the three targets in binary systems J1317-0157, J0024-7204H and J0024-7204S; see Tables~\ref{tab:doublets}-~\ref{tab:triplets}. 

Following the discussion in~\cite{Leaci:2016oja,PhysRevD.91.102003}, the binary Doppler correction can be expressed in terms of five binary orbital parameters: $a_p$, the projected semi-major axis of the orbit along the line of sight normalised to $c$; $P_{\rm orb}$, the orbital period; $e$, the orbit eccentricity; $\omega$, the argument of periastron; and $t_p$, the time of periapse's passage. In the regime of low eccentricities, $e\ll1$, $e,\,\omega,\,t_p$ are not directly provided in the ephemerides due to strong correlations among the parameters. Instead, astronomers replace those with $t_{\rm asc} = t_p - 2\pi\omega/P$, $\eta = e\sin\omega$, and $\kappa = e\cos\omega$. The parameters used in the analysis are taken from Ref.~\cite{Swiggum:2022xlb} for J1317-0157, while the other two ephemerides are taken from the ATNF catalog~\cite{pulsarlist,Manchester:2004bp} and are reported in Table~\ref{tab:errors}.

\begin{table*}[!hpt]
    \centering
    \caption{Five keplerian parameters and their reported error for the 3 pulsars in a binary system analyzed in this work. For the targets J0024-7204H and J1317-0157, we report the low eccentricity approximation parameters (starred values below). Values are taken from Ref.~\cite{Swiggum:2022xlb} for J1317-0157, while the other two ephemerides are taken from the ATNF catalog~\cite{pulsarlist,Manchester:2004bp}.}
    \begin{tabular}{cccccc}
    \hline
    \hline
        Pulsar name & $P$~[day] & $a_p$ [ls]& $e/\eta^*$ & $\omega$ [$^\circ$] /$\kappa^*$ & $t_p/t_{\rm asc}^*$ [MJD]\\
        \hline
         J0024-7204H &  $2.35769689(1)$ & $2.152813(2)$ & $0.070558(1)$ & $111.318(4)$ & $51602.186289(7)$\\
         J0024-7204S &  $1.201724200(3)$ & $0.7662686(8)$ & $9.1(3)\cdot10^{-5*}$ & $3.9(2)\cdot10^{-4*}$ & $51600.6250241(7)^*$\\
         J1317-0157 &  $0.089128297(2)$ & $0.027795(4)$ & $5(2)\cdot10^{-4*}$ & $0(2)\cdot10^{-4*}$ & $57909.041863(5)^*$\\
         \hline
    \end{tabular}
    \label{tab:errors}
\end{table*}

\begin{table*}[!hpt]
\renewcommand{\arraystretch}{1.2}
    \centering
    \caption{Results from the semicoherent method for three targets in binary systems, following Ref.~\cite{Mirasola:2024kll}. We report the coherence time, with starred values indicating that the analysis has been done with $\Tfft=1/5 T_{\rm sid}$, the $CR$ showing that no outliers have been identified ({\it i.e.}, $CR<CR_{\rm thr}$), and the estimated strain ULs.}
    \begin{tabular}{ccccc}
    \hline
    \hline
        Pulsar name & $\Tfft$ [$T_{\rm sid}$] & $CR_{\rm L1}$ & $CR_{\rm H1}$ & $h_{\rm UL}$\\
        \hline
         J0024-7204H &  2.96$^*$ & 0.20 & -0.20 & $8.9\times10^{-26}$ \\
         J0024-7204S &  1.84$^*$ & 1.01 & 1.15 & $1.1\times10^{-25}$ \\
         J1317-0157  &  0.17 & 2.08 & 0.43 & $1.7\times10^{-25}$ 
         \\\hline
    \end{tabular}
    \label{tab:semi-coh_bin_res}
\end{table*}

We observe that the uncertainties on the binary parameters of J0024-7204H (see Table~\ref{tab:errors}) are within the maxima offsets of these parameters for which the resampling method has been tested and proven to successfully recover the five peaks [see Table 3 of~\cite{Singhal:2019dfn}].
On the other hand, the error bars for J0024-7204S and J1317-0157 are larger than these bounds. In fact, converting the low eccentricity parameters and their uncertainties given in Table~\ref{tab:errors} in terms of the nominal 5 Keplerian parameters [to compare with the work in~\cite{Singhal:2019dfn}], we had to explore 85 templates in the $\omega\times t_p$ space and 91 $\omega$ values to cover J0024-7204S and J1317-0157 uncertainties, respectively.

\section{Semicoherent method for targets in binary system\label{app:semi-coh_method_bin}}

Additionally to the method presented in Sec.~\ref{sec:resamp_pipeline}, we apply the search presented in~\cite{Mirasola:2024kll} for the three targets in binary systems (see Appendix~\ref{app:resamp_grid} and Table~\ref{tab:errors}).
The outcome of this search is not included in the main text as its results are less sensitive than the method presented in Sec.~\ref{sec:resamp_pipeline}.

Here, data in the time domain from the \textit{Band Sampled Data} framework~\cite{Piccinni:2018akm} are split into segments of a given duration, known as coherence time and indicated with $\Tfft$. The $\Tfft$ is calculated for each target given their parameters and takes into account the variation of the orbital modulation within their uncertainties.

Data are initially Doppler corrected and then split into segments of length $\Tfft$. Each segment is Fourier-transformed and whitened using the average spectrum along the data stream~\cite{Piccinni:2018akm}. After the equalization procedure, we select local maxima in the frequency domain above a threshold of $\theta=2.5$~\cite{Astone:2014esa}. 
Then, we construct the histogram of selected frequencies and consider as a potential candidate the loudest peak around the expected frequency, in the range $f_{\rm GW}\pm 1.5\> \delta f$, where $\delta f$ is the Fourier transform frequency resolution given by $1/\Tfft$.

To decide whether the candidate has to be considered an outlier, our detection statistic is again the $CR$ as defined in Eq.~\eqref{eq:CR}. In this case, however, for each candidate, $\mathcal{S}$ is replaced by the number of times that a specific bin has been selected. While $\mu$ and $\sigma$ are the average and standard deviation, respectively, of the noise counts calculated outside the region from which the loudest peak is selected. Similarly to the other analyses, we set a threshold of FAP=1\% that corresponds to $CR_{\rm thr}\approx2.79$ (see Appendix~E of~\cite{Mirasola:2024kll}).
Here, different detector data are analyzed independently and their results are compared for consistency checks. Due to a current software limitation, the study is bounded to $\Tfft<T_{\rm sid}/5\sim 1.7\cdot10^4$~s, where $T_{\rm sid}$ is the sidereal day.

Our search did not highlight any significant outlier, as shown by the $CR$ values reported in Table~\ref{tab:semi-coh_bin_res}. We therefore set ULs at the 95\% CL through fake signals injection in real data and report them in Table~\ref{tab:semi-coh_bin_res}; the resulting ULs are a factor of 1.2-5 higher with respect to the ULs obtained with the resampling pipeline for the same sources.

\section{Source parameters}~\label{app:source_params}

In this section, we report the measured values of the ephemerides used for all 34 sources analyzed in this work and an additional 10 sources are used to constrain dark photon signals using previous results. The sources are divided into the three categories of ``$\dot{f} > 0$'', ``frequency-doublets'', and ``frequency-triplets''  (see Sec.~\ref{sec:targets} for more details on the selection criteria), reported in Tables~\ref{tab:doublets},~\ref{tab:triplets}, and~\ref{tab:pos_fdot}, respectively. 
In each table, we indicate the source name, position in the sky using the equatorial coordinates ($\alpha, \delta$), the rotational frequency ($f_{\rm EM}$), and its time derivative ($\dot{f}_{\rm EM}$). Values are taken from Ref.~\cite{Swiggum:2022xlb} for two pulsars (J1122-3546 and J1317-0157), while all the others are from the ATNF catalog~\cite{pulsarlist,Manchester:2004bp}. We additionally report the maximum theoretically expected strain, $h_{\rm max}$, that a dark photon
cloud could produce, under the optimistic assumption that it is as young as $10^3$ yr~\cite{Siemonsen:2022ivj}. For sources with $\dot{f}_{\rm rot}$, we additionally include a minimum  theoretically expected strain, $h_{\rm min}$, obtained by assuming that spin-up is entirely due to the superradiance cloud decay through GW emission~\cite{Siemonsen:2022ivj}. The pipeline used to analyze each source, together with the obtained strain ULs are reported in Table~\ref{tab:results} in the main text.

Note that three sources in the categories ``frequency-doublets'' and ``frequency-triplets'' have $\dot{f}_{\rm EM} > 0$ (J1748-3009, J0024-7204S, and J0024-7204H). These sources are known to be in a binary system, and hence the positive spin frequency derivatives are more likely linked to the source’s acceleration in the binary orbit; therefore we do not include them in the set of potentially spinning-up sources ``$\dot{f} > 0$''.

\setlength{\tabcolsep}{5pt}
\renewcommand{\arraystretch}{1.6}
\LTcapwidth=\textwidth
\begin{longtable*}{ccccccc}
\caption{Parameters of the targets with $\dot{f}>0$. Frequencies are propagated to the epoch 58700 MJD. Targets with the superscript $D$ happen to also be in a frequency-doublet. Values are taken from the ATNF catalog~\cite{pulsarlist,Manchester:2004bp}. }\\
\hline \hline \rule{0pt}{3ex}  
Pulsar name & $\alpha$ & $\delta$  & $f_{\text{EM}}$ & $\dot{f}_{\text{EM}}$ & $h_{\text{min}}$ & $h_{\text{max}}$\\ 
(J2000)  & (hh:mm:ss) & (hh:mm:ss) & (Hz) & (Hz/s) & $\times 10^{-25}$ & $\times 10^{-25}$ 
\label{tab:pos_fdot}
\\\hline
\endfirsthead
J1757-2745 & 17:57:54.7826(1) & -27:45:40.16(2) & $56.53801333052(5)$ & $4.17(5) \times 10^{-17}$ & 15.1 & 19.6\\
J1641+3627A & 16:41:40.87019(5) & +36:27:14.9788(4) & $96.362234567(4)$ & $6.75(3) \times 10^{-16}$ & 25.0 & 29.2\\
J1748-2446C & 17:48:04.54(1) & -24:46:36(4) & $118.53825306(4)$ & $8.52(6) \times 10^{-15}$ & 61.0 & 63.1\\
J1910-5959B & 19:10:52.0556(5) & -59:59:00.861(6) & $119.6487328450(5)$ & $1.13154(8) \times 10^{-14}$ & 119.0 & 119.3\\
J1801-0857A & 18:01:50.6097(1) & -08:57:31.853(6) & $139.360884729(3)$ & $9.9124(8) \times 10^{-15}$ & 31.2 & 32.2\\
J1748-2021C & 17:48:51.17320(1) & -20:21:53.81(4) & $160.592709908(2)$ & $1.543(4) \times 10^{-15}$ & 11.7 & 13.4\\
J0024-7204C & 00:23:50.3546(1) & -72:04:31.5048(4) & $173.70821896596(4)$ & $1.50421(6) \times 10^{-15}$ & 17.9 & 20.7\\
J1911+0101B & 19:11:12.5725(4) & +01:01:50.44(2) & $185.724277202(5)$ & $7(1) \times 10^{-17}$ & 2.1 & 2.8\\
J0024-7204D & 00:24:13.88092(6) & -72:04:43.8524(2) & $186.65166985673(2)$ & $1.1922(3) \times 10^{-16}$ & 4.4 & 5.6\\
J0024-7204Z$^D$ & 00:24:06.041(2) & -72:05:01.480(6) & $219.565606035(2)$ & $2.19(3) \times 10^{-16}$ & 4.3 & 5.5\\
J0024-7204L & 00:24:03.7721(3) & -72:04:56.923(2) & $230.0877462914(1)$ & $6.4611(2) \times 10^{-15}$ & 21.5 & 23.3\\
J0024-7204G$^D$ & 00:24:07.9603(1) & -72:04:39.7030(5) & $247.50152509638(6)$ & $2.5825(1) \times 10^{-15}$ & 11.8 & 13.4\\
J1801-0857C$^D$ & 18:01:50.73731(7) & -08:57:32.699(3) & $267.47267494(3)$ & $4.573(9) \times 10^{-15}$ & 6.0 & 6.6\\
J0024-7204M$^D$ & 00:23:54.4899(3) & -72:05:30.756(2) & $271.9872287887(2)$ & $2.8421(4) \times 10^{-15}$ & 10.3 & 11.7\\
J1836-2354B & 18:36:24.351(3) & -23:54:28.7(7) & $309.379715192(2)$ & $4.6(6) \times 10^{-17}$ & 1.5 & 1.6\\
J0024-7204N & 00:24:09.1880(2) & -72:04:28.8907(7) & $327.4443186174(1)$ & $2.3435(2) \times 10^{-15}$ & 6.5 & 6.9
\\\hline   
\end{longtable*}

\setlength{\tabcolsep}{5pt}
\renewcommand{\arraystretch}{1.6}
\LTcapwidth=\textwidth
\begin{longtable*}{cccccc}
\caption{Parameters of the targets in doublets. Frequencies are propagated to the epoch 58700 MJD. Additional targets in this group are listed in Table~\ref{tab:pos_fdot} with the superscript $D$. Note that targets might be missing the companion, due to the selections described in Sec.~\ref{subsec:targets_and_methods}. Values are taken from the ATNF catalog~\cite{pulsarlist,Manchester:2004bp} for all the targets but J1122-3546 and J1317-0157, where we considered the ephemerides reported in Ref.~\cite{Swiggum:2022xlb}.}\\
\hline \hline \rule{0pt}{3ex}  
Pulsar name & $\alpha$ & $\delta$  & $f_{\text{EM}}$ & $\dot{f}_{\text{EM}}$ & $h_{\text{max}}$\\ 
(J2000)  & (hh:mm:ss) & (hh:mm:ss) & (Hz) & (Hz/s) & $\times 10^{-25}$ 
\label{tab:doublets}
\\\hline   
\endfirsthead
J1748-3009 & 17:48:24(0) & -30:09:11(0) & $103.263527(0)$ & $3.77(0)\times 10^{-16}$  & 16.1\\
J0921-5202 & 9:21:00(0) & -52:02:00(0) & $103.30777(0)$ & $-2.209(0)\times 10^{-16}$  & 229.7\\
J1122-3546 & 11:22:17.24(0) & -35:46:31.2(0) & $127.582438285(0)$ & $-2.48(0) \times 10^{-16}$ & 124.5\\
J1546-5925 & 15:46:29.3456(5) & -59:25:44.871(3) & $128.2589171003(7)$ & $-2.65(5) \times 10^{-16}$ & 20.4\\
J1551-0658 & 15:51:07.215(4) & -06:58:06.5(6) & $140.9690(0)$ & $-3.99(0)\times 10^{-16}$ & 61.7\\
J1748-2446T & 17:48:04.8(0) & -24:46:45(0) & $141.145053(0)$ & ... & 11.8\\
J0514-4002N & 05:14:06.7(1) & -40:02:47(1) & $179.600927(0)$ & ... & 7.1\\
J0514-4002C & 05:14:06.69(4) & -40:02:49.4(3) & $179.700978(0)$ & ... & 7.1\\
J1824-2452E & 18:24:32.81(0) & -24:52:11.20(0) & $184.501845(0)$ & ... & 14.9\\
J1838-0022g & 18:38:24(6) & +00:22:00(2) & $196.463654(0)$ & ... & 20.1\\
J1748-2446ac & 17:48:04.8(0) & -24:46:45(0) & $196.582994(0)$ & ... & 11.8\\
J1940+26 & 19:40:13(8) & +26:01:00(2) & $207.74778(7)$ & ... & 10.3\\
J1904+0836g & 19:04:35(6) & +08:36:00(2) & $225.225225(0)$ & ... & 27.9\\
J1953+1844g & 19:53:44(7) & +18:44:00(2) & $225.225225(0)$ & ... & 18.8\\
J1326-4728E & 13:26:44(0) & -47:30:00(0) & $237.658566471(2)$ & ... & 15.7\\
J0125-2327 & 1:25:10(0) & -23:27:26(0) & $272.0810887(0)$ & $-1.3567(0)\times 10^{-15}$ & 93.6\\
J1844+0028g & 18:44:36(6) & +00:28:48(12) & $280.112045(0)$ & ... & 17.9\\
J1317-0157 & 13:17:40.45(0) & -1:57:30.1(0) & $343.8500325(2)$ & $-6.5(4)\times 10^{-16}$ & 3.3\\
J0418+6635 & 04:18:47.9811(2) & +66:35:24.726(3) & $343.62080(1)$ & $-1.613(1) \times 10^{-15}$ & 37.0\\
J0024-7204S & 00:24:03.9794(1) & -72:04:42.3530(4) & $353.30621861542(6)$ & $1.50466(1)\times 10^{-14}$ & 17.4\\
J1308-23 & 13:08:00(0) & -23:00:00(0) & $353.4(1)$ & ... & 47.7
\\\hline   
\end{longtable*}

\setlength{\tabcolsep}{5pt}
\renewcommand{\arraystretch}{1.6}
\begin{longtable*}{cccccc}
\caption{Parameters of the targets in triplets. Frequencies are propagated to the epoch 58700 MJD. Note that targets might be missing the companion, due to the selections described in Sec.~\ref{subsec:targets_and_methods}. Values are taken from the ATNF catalog~\cite{pulsarlist,Manchester:2004bp} for all the targets.}\\
\hline \hline \rule{0pt}{3ex}  
Pulsar name & $\alpha$ & $\delta$  & $f_{\text{EM}}$ & $\dot{f}_{\text{EM}}$  & $h_{\text{max}}$\\ 
(J2000)  & (hh:mm:ss) & (hh:mm:ss) & (Hz) & (Hz/s) & $\times 10^{-25}$ 
\label{tab:triplets}
\\\hline  
\endfirsthead
J1342+2822F & 13:42:11.6(0) & +28:22:38(0) & $227.3(0)$ & ... & 8.0\\
J1803-3002B & 18:03:34.0(0) & -30:02:02(0) & $227.427792(0)$ & ... & 10.6\\
J1823-3021E & 18:23:38.93(0) & -30:22:22.3(0) & $227.583068(0)$ & ... & 10.3\\
J0024-7204H & 00:24:06.7032(2) & -72:04:06.8067(6) & $311.49341795311(6)$ & $1.775(1)\times10^{-16}$ & 17.4\\
J1930+1403g & 19:30:18(7) & +14:03(2) & $311.526480(0)$ & ... & 17.2\\
J1624-39 & 16:24:25(0) & -39:51:00(0) & $337.837838(0)$ & ... & 31.1\\
J2045-68 & 20:45:17(0) & -68:36:20(0) & $337.837838(0)$ & ... & 64.2
\\\hline   
\end{longtable*}

\begin{widetext}
\justifying

\section{Expected number of observable events\label{app:events_derivation}}

\subsection{Strain density distribution} \label{app:strain_dist_disk}

To obtain the final expression for the strain density distribution of a signal of strain $h_0$ from Galactic disk BHs given in Sec.~\ref{sec:CWh_dist}, we start from the definition given in Eq.~\eqref{eq:dfdlogh},
\begin{align}\label{eq:dfdlogh_app}
    \frac{\dd n_{h}}{\dd h_0}(h_0\ |\ \dpm, \varepsilon) & = \frac{1}{\mathcal{N}\chi_{\rm max}t_{\rm max}} \int \dd M \frac{M_{\mathrm{min}}^{1.35}}{M^{2.35}} \int \dd \chi \int \dd \vec{r} \frac{\dd n_{\rm disk}}{\dd \vec{r}}\int \dd t\ \delta[\bar{h}_0(M, \chi, \tau_{\rm obs}, d\ |\ \dpm) - h] \Theta_{\dot{f}} \Theta_{\rm EM} \Theta_{\rm pl} \Theta_{\rm ev}. 
\end{align}
The strain dependence with respect to time and distance can be written out explicitly from Eq.~\eqref{eq:strainevo} as
\begin{equation}
    \bar{h}_0(M, \chi, \tau_{\rm obs}, d) = \frac{h_{r_d}(M, \chi)}{1 + \tau_{\rm obs}/\tau_{\rm GW}(M,\chi)}\frac{r_d}{d},
\end{equation}
where $\tau_{\rm GW}$ is the characteristic timescale for GW emission [see Eq.~\eqref{eq:tgw}] and $h_{r_d}(M, \chi) \equiv h^{s}_0(M, \chi, d=r_d)$ is the strain emitted at cloud saturation evaluated at a distance equal to the disk's characteristic size $r_d$, for later convenience in the overall normalization. The $\delta$-function in Eq.~\eqref{eq:dfdlogh_app} is satisfied for $\bar{\tau} = $ $\tau_{\rm SR} + d/c + \tau_{\rm GW} \left[h_{r_d} r_d/(h_0 d) - 1\right]$ $ \equiv \tau_{\rm SR} + d/c + \bar{\tau}_{\rm obs}$ and can be used to integrate over the BH's age, to get 
\begin{equation}\label{eq:dfdlogh_2}
      \frac{\dd n_{h}}{\dd \log h_0} = \frac{1}{\mathcal{N}} \int \dd M \frac{M_{\mathrm{min}}^{1.35}}{M^{2.35}} \int \dd \chi\ \frac{1}{\chi_{\rm max}}\frac{h_{r_d} (M, \chi)}{h_0} \frac{\tau_{\rm GW}(M, \chi)}{\tau_{\rm max}} \Theta_{\rm pl}  \int \dd \vec{r} \frac{\dd n_{\rm disk}}{\dd V}\ \frac{r_d }{d} \Theta_{\dot{f}} \Theta_{\rm EM} \Theta_{\rm ev} ,
\end{equation}
where we have dropped the dependence on the dark photon parameters for simplicity and the boundary conditions of the volume integral are determined by the boundary conditions of the age integral, requiring
\begin{equation}
    {\rm max}\lbrace \tau_{\rm min}, \tau_{\rm SR}+d/c \rbrace < \bar{\tau} < \tau_{\rm max}.
\end{equation}
Given the cylindrical symmetry of the BH distribution in Eq.~\eqref{eq:spatial_disk}, the location of the observer can be conveniently chosen along one of the horizontal axis,  {\it i.e.} $\lbrace \rho, \phi, z \rbrace = \lbrace r_{\rm obs}, 0, 0 \rbrace$ in cylindrical coordinates, with $r_{\rm obs} \simeq 8$ kpc. Writing out the disk spatial distribution explicitly, we get
\begin{equation}\label{eq:dfdlogh_3}
\frac{\dd n_{h}}{\dd \log h_0} = \frac{x_{\rm disk }N_{\rm BH}}{4\pi \mathcal{N}} \int \dd M \frac{M_{\mathrm{min}}^{1.35}}{M^{2.35}} \int \dd \chi\ \frac{1}{\chi_{\rm max}}\frac{h_{r_d}}{h_0} \frac{\tau_{\rm GW}}{\tau_{\rm max}}\Theta_{\rm pl}  \int \dd \Tilde{z} \int \dd \varphi \int \dd \Tilde{\rho}\frac{\Tilde{\rho}\  e^{-\Tilde{\rho}}\ \Theta_{\dot{f}} \Theta_{\rm EM} \Theta_{\rm ev}}{\sqrt{\Tilde{\rho}^2 + \Tilde{r}_{\rm obs}^2 + (z_{\rm max}/r_d)^2 \Tilde{z}^2 - 2 \Tilde{\rho}\ \Tilde{r}_{\rm obs} \cos \varphi}},
\end{equation}
where we have introduced $\Tilde{z} \equiv z/z_{\rm max}$, $\Tilde{\rho} \equiv \rho/r_d$, and $\Tilde{r}_{\rm obs} = r_{\rm obs}/r_d$. If we now take the thin disk approximation and assume $d \simeq \sqrt{\rho^2 + r_{\rm obs}^2 - 2 \rho\ r_{\rm obs} \cos \varphi}$, the above expression simplifies to Eq.~\eqref{eq:dfdlogh_disk} from the main text. The boundaries of the volume integral in Eq.~\eqref{eq:dfdlogh_disk} have to satisfy the conditions
\begin{align}
    d_{-}^{\rm max} & < \sqrt{\rho^2 + r_{\rm obs}^2 - 2 \rho\ r_{\rm obs} \cos \varphi} < {\rm min} \left\lbrace d_{+}^{\rm max}, \frac{h_{r_d}}{h_0}r_d, \frac{\tau_{\rm max} - \tau_{\rm SR}}{c} \right\rbrace \nonumber \\
    d_{\pm}^{\rm max} & \equiv c\frac{\tau_{\rm max} - \tau_{\rm SR} + \tau_{\rm GW} \pm \sqrt{(\tau_{\rm max} - \tau_{\rm SR} + \tau_{\rm GW})^2 - 4 \tau_{\rm GW} (r_d /c)( h_{r_d}/h_0)}}{2}.
\end{align}
For the $\Theta_{\dot{f}}$ function--see Sec.~\ref{sec:CWh_dist}--the explicit time evolution of the frequency derivative is determined by $\dot{f}_{\rm GW} \propto \dot{M}_c(\tau)$--see Sec.~\ref{sec:dp_superrad}--and is given by 
\begin{align}
    \dot{f}_{\rm GW} (t) = \frac{\dot{f}^s_{\rm GW}}{\left(1+\tau/\tau_{\rm GW}\right)^2}.
\end{align}
In Fig.~\ref{fig:dndlogh_fr7}, we show the strain distribution from Eq.~\eqref{eq:dfdlogh_disk} and the total expected number of events from Eq.~\eqref{eq:nev_aboveUL} similarly to the results shown in Fig.~\ref{fig:dndlogh} in the main text, but fixing the fraction of radio emission of the dark photon cloud to $f_r = 10^{-7}$, instead of $f_r = 10^{-5}$.
\begin{figure*}[t]
    \centering 
\includegraphics[width=0.48\textwidth]{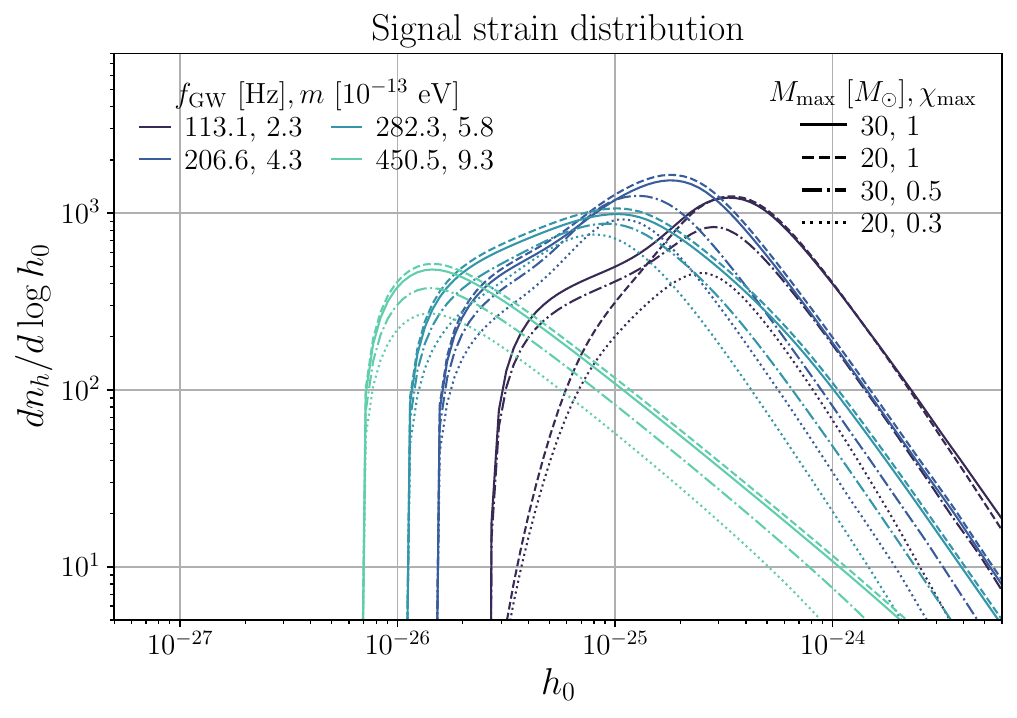}
\includegraphics[width=0.48\textwidth]{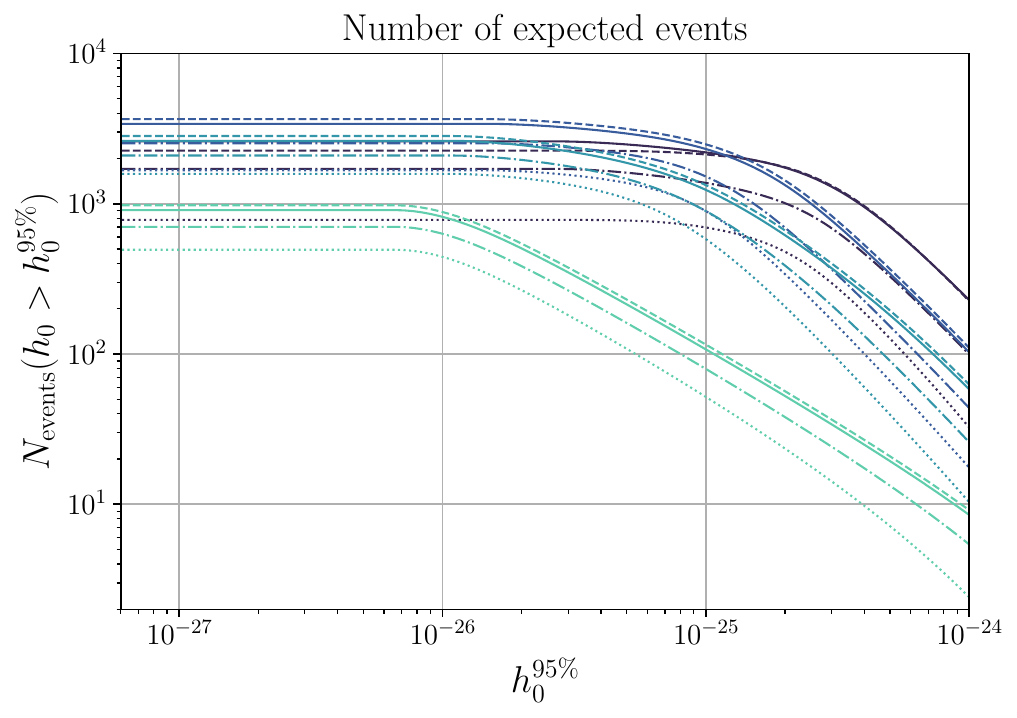}
    \caption{Same as Fig.~\ref{fig:dndlogh}, but fixing the fraction of EM emission into radio to $f_r = 10^{-7}$. We computed the result also for $f_r = 10^{-3}$ and curves are unchanged from Fig.~\ref{fig:dndlogh}.
    }\label{fig:dndlogh_fr7}
\end{figure*}

\subsection{Expected events for other sources}\label{app:Nevents_all}

In this section we show the number of expected observable events for the sources that have not being analyzed with the narrow-band method, to complement the results shown in Sec.~\ref{sec:CWh_dist} (excluding the three sources that are known to be in a binary system, since we are only considering the population of isolated BHs). The results are shown in Fig.~\ref{fig:Nev_targets_all}.

\begin{figure*}[t]
    \centering \includegraphics[width=0.48\textwidth]{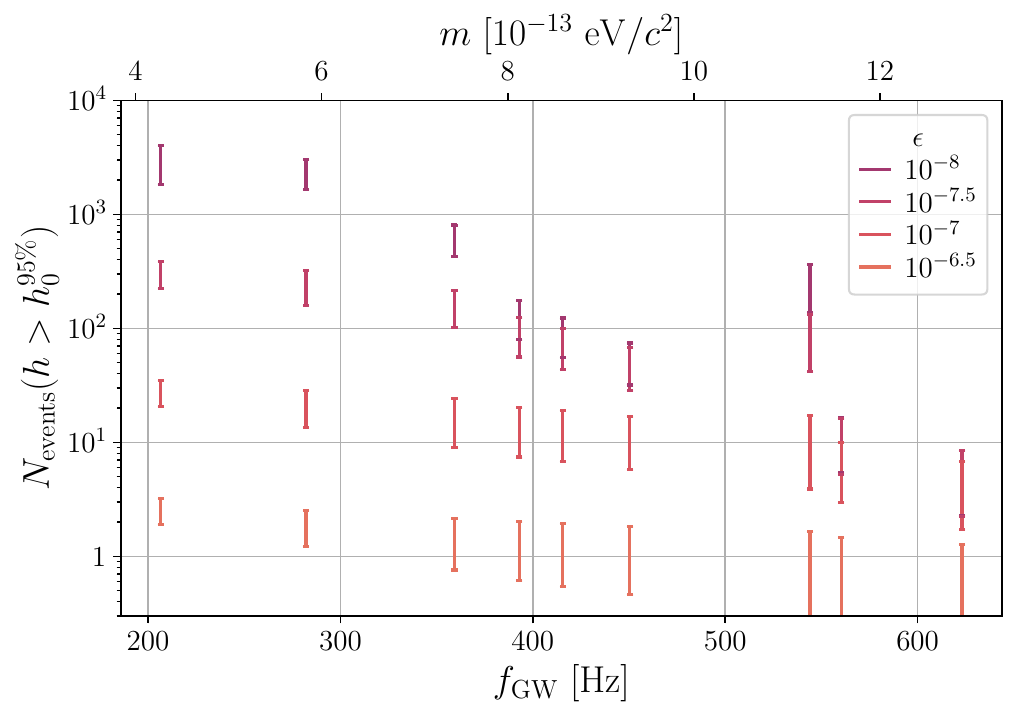}
    \includegraphics[width=0.48\textwidth]{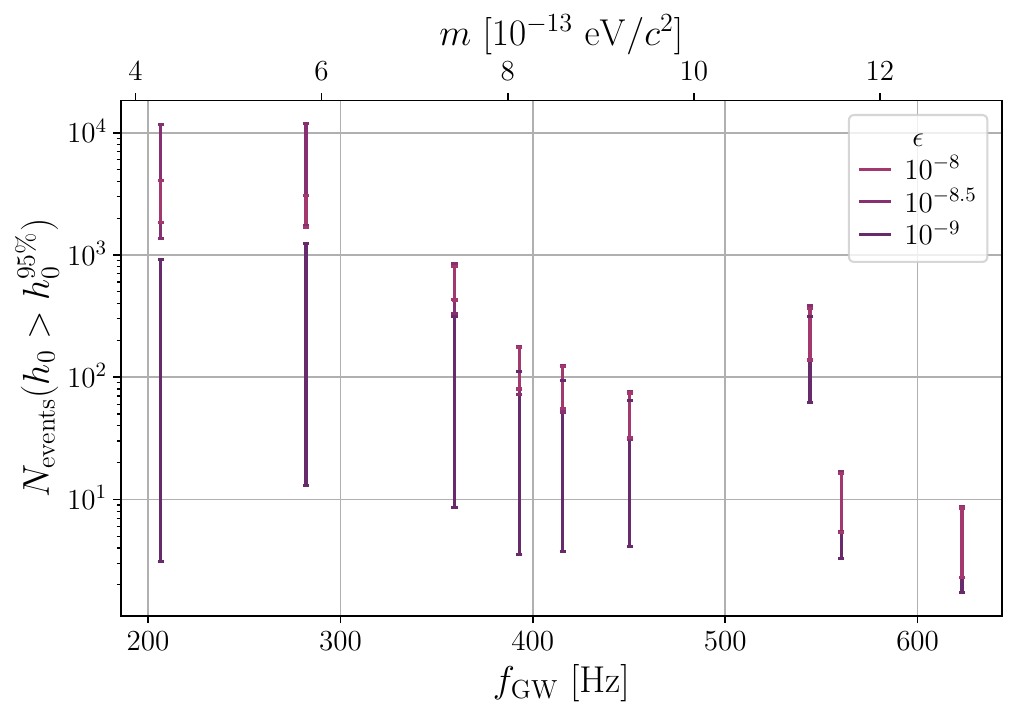}
    \caption{Same as Fig.~\ref{fig:Nev_targets}, but for the sources in Table~\ref{tab:results} that were analyzed with a pipeline different from the narrow-band method.
    }\label{fig:Nev_targets_all}
\end{figure*}

\subsection{Radio flux density distribution} \label{app:flux_dist_disk}

Similarly to the derivation presented in Appendix~\ref{app:strain_dist_disk}, to obtain the radio luminosity function of dark photon superradiance clouds from Galactic disk BHs, we start from the density distribution of a signal with radio flux $F_r$
\begin{align}\label{eq:dfdlogF_0}
    \frac{\dd n_{F_r}}{\dd F_r}(F_r\ |\ \dpm, \varepsilon) & = \int \dd \vec{r} \dd M \dd \chi \dd \tau \frac{\dd n_{\rm BH}}{\dd \vec{r}\, \dd M \dd \chi \dd \tau}\ \delta\left[\bar{F}_r(M, \chi, \tau_{\rm obs}, d\ |\ \dpm, \varepsilon) - F_r\right] \Theta_{\rm pl} \Theta_{\rm ev} \nonumber \\
    & = \frac{1}{\mathcal{N}\chi_{\rm max}\tau_{\rm max}} \int \dd M \frac{M_{\mathrm{min}}^{1.35}}{M^{2.35}} \int \dd \chi \Theta_{\rm pl} \int \dd \vec{r} \frac{\dd n_{\rm disk}}{\dd \vec{r}}\int \dd t\ \delta\left[\bar{F}_r(M, \chi, \tau_{\rm obs}, d\ |\ \dpm, \varepsilon) - F_r\right] \Theta_{\rm ev}, 
\end{align}
where the BHs differential distribution is given in Eq.~\eqref{eq:BH_dist} and, compared to the CW strain distribution from Eq.~\eqref{eq:dfdlogh}, we do not need to include $\Theta_{\rm EM}$ and $\Theta_{\dot{f}}$.\footnote{Radio surveys would loose sensitivity for pulsating signals that are rapidly varying over time, {\it i.e.~}with a large spin-up rate. However, we have already seen in Sec.~\ref{sec:CWh_dist} that most of the observable systems have a very small spin-up rate. Therefore we neglect the condition on $\dot{f}$ for simplicity.} The radio flux dependence with respect to time, distance, and $\varepsilon$ can be written out explicitly from Eq.~\eqref{eq:radio_lum} as
\begin{equation}
    \bar{F}_{\rm r}(M, \chi, \tau_{\rm obs}, d\ |\ \dpm, \varepsilon) = f_r\frac{\varepsilon^2 L^s(M, \chi|\ \dpm, \varepsilon=1)}{1 + \tau_{\rm obs}/\tau_{\rm GW}(M,\chi)} \frac{1}{4\pi d^2},
\end{equation}
where $\tau_{\rm GW}$ is the characteristic timescale for GW emission [see Eq.~\eqref{eq:tgw}], $L^s$ is the EM luminosity at cloud saturation, given by taking $M_c = M_c^s$ in Eq.~\eqref{eq:lum}, and $f_r$ is the fraction going into radio emission. The $\delta$-function in Eq.~\eqref{eq:dfdlogF_0} is satisfied for $\bar{\tau} \equiv \tau_{\rm SR} + d/c + \tau_{\rm GW} \left[f_r \varepsilon^2 L^s/(4\pi d^2 F_r)- 1\right] \equiv \tau_{\rm SR} + d/c + \bar{\tau}_{\rm obs}$ and can be used to integrate over time to get 
\begin{equation}\label{eq:dfdlogF_1}
      \frac{\dd n_{F_r}}{\dd \log F_r} = \frac{1}{\mathcal{N}} \int \dd M \frac{M_{\mathrm{min}}^{1.35}}{M^{2.35}} \int \dd \chi\ \frac{1}{\chi_{\rm max}}\frac{f_r \varepsilon^2 L^s(M, \chi)}{4\pi r_d^2 F_r} \frac{\tau_{\rm GW}(M, \chi)}{\tau_{\rm max}} \Theta_{\rm pl}  \int \dd \vec{r} \frac{\dd n_{\rm disk}}{\dd \vec{r}}\, \frac{r_d^2 }{d^2} \,  \Theta_{\rm ev} ,
\end{equation}
where we have dropped the dependence on the dark photon parameters for simplicity and the boundary conditions of the volume integral are determined by the boundary conditions of the time integral, requiring
\begin{equation}
    {\rm max}\lbrace \tau_{\rm min}, \tau_{\rm SR}+d/c \rbrace < \bar{\tau} < \tau_{\rm max}.
\end{equation}
Given the cylindrical symmetry of the BH distribution in Eq.~\eqref{eq:spatial_disk}, the location of the observer can be conveniently chosen along one of the horizontal axis,  {\it i.e.}, $\lbrace \rho, \phi, z \rbrace = \lbrace r_{\rm obs}, 0, 0 \rbrace$ in cylindrical coordinates, with $r_{\rm obs} \simeq 8$ kpc. Writing out the disk spatial distribution explicitly, we get
    \begin{equation}\label{eq:dfdlogF_2}
      \frac{\dd n_{F_r}}{\dd \log F_r} = \frac{x_{\rm disk} \,N_{\rm BH}}{4\pi\mathcal{N}} \int \dd M \frac{M_{\mathrm{min}}^{1.35}}{M^{2.35}} \int \dd \chi\ \frac{1}{\chi_{\rm max}}\frac{f_r \varepsilon^2 L^s}{4\pi r_d^2 F_r} \frac{\tau_{\rm GW}}{\tau_{\rm max}} \Theta_{\rm pl} \int \dd \Tilde{z} \int \dd \varphi \int \dd \Tilde{\rho}  \frac{\Tilde{\rho}  e^{-\Tilde{\rho}}\ \Theta_{\rm ev}}{\Tilde{\rho}^2 + \Tilde{r}_{\rm obs}^2 + (z_{\rm max}/r_d)^2 \Tilde{z}^2 - 2 \Tilde{\rho}\ \Tilde{r}_{\rm obs} \cos \varphi}   ,
\end{equation}
where we have introduced $\Tilde{z} \equiv z/z_{\rm max}$, $\Tilde{\rho} \equiv \rho/r_d$, and $\Tilde{r}_{\rm obs} = r_{\rm obs}/r_d$. If we now take the thin disk approximation and assume $d \simeq \sqrt{\rho^2 + r_{\rm obs}^2 - 2 \rho\ r_{\rm obs} \cos \varphi}$, the above expression simplifies to Eq.~\eqref{eq:dfdlogF_disk} from the main text. The boundary conditions of the volume integral over $\varphi$ and $\rho$ in Eq.~\eqref{eq:dfdlogF_disk} have to satisfy 
\begin{align}
    & d^2 < \frac{f_r \varepsilon^2 L^s}{4\pi F_r} \\
    & d^3 + c(\tau_{\rm SR} - \tau_{\rm GW}-\tau_{\rm max}) d^2 +  \frac{f_r \varepsilon^2 L^s}{4\pi F_r} c \tau_{\rm GW}  < 0,
  \end{align}
where $d^2 = \rho^2 + r_{\rm obs}^2 - 2 \rho\ r_{\rm obs} \cos \varphi$.

\end{widetext}
\setcounter{tocdepth}{2}

\bibliography{bib.bib}

\end{document}